\documentclass[aps,prl,twocolumn,superscriptaddress,longbibliography,epsfig,floats,preprintnumbers,floatfix]{revtex4-1}
\usepackage{graphicx}
\usepackage[latin2]{inputenc}
\usepackage{amsmath}
\usepackage{amssymb}
\usepackage{bm}
\usepackage{color}
\usepackage{sidecap}

\definecolor{NatureBlue}{rgb}{0.012,0.3,0.63}
\definecolor{Red}{rgb}{0.81,0.09,0.13}

\newcommand{\Red}{\textcolor{Red}}

\newcommand{\fspVec}{{\bm F}\!_{_\text{sp}}}
\newcommand{\fsp}{F\!_{_\text{sp}}}
\newcommand{\fmax}{F\!_{_\text{max}}}
\newcommand{\fnc}{F\!_{\theta{=}0}}
\newcommand{\topt}{\theta_{_\text{opt}}}
\newcommand{\funit}{k\!_{_B}\!T\!{/}\sigma}
\newcommand{\kbt}{k\!_{_B}T}

\newcommand{\Vint}{{\hat{\text{\bf e}}}}

\begin{document}
\title{\Red{Optimal chirality enhances long-range fluctuation-induced interactions in active fluids}}
\author{Hashem Fatemi}
\affiliation{School of Quantum Physics and Matter, Institute for Research in Fundamental Sciences 
(IPM), Tehran 19538-33511, Iran}
\author{Hamidreza Khalilian}
\affiliation{School of Quantum Physics and Matter, Institute for Research in Fundamental Sciences 
(IPM), Tehran 19538-33511, Iran}
\author{Jalal Sarabadani$^*$}
\affiliation{School of Quantum Physics and Matter, Institute for Research in Fundamental Sciences 
(IPM), Tehran 19538-33511, Iran}
\author{Reza Shaebani$^*$}
\affiliation{Department of Theoretical Physics and Center for Biophysics, 
Saarland University, 66123 Saarbr\"ucken, Germany}

\begin{abstract}
Understanding interactions between chiral active particles--- self-propelling and self-rotating 
entities--- is crucial for uncovering how chiral active matter self-organizes into dynamic 
structures. Although fluctuation-induced forces in nonequilibrium active systems can drive 
structure formation, the role of chirality remains largely unexplored. We investigate effective 
fluctuation-induced forces between intruders immersed in chiral active fluids and reveal that 
the impact of chirality depends sensitively on particle shape. For circular particles, increasing 
the self-rotation to self-propulsion ratio suppresses the interaction, reflecting a transition 
from rotating flocks to localized spinners. Contrarily, a striking collective behavior emerges 
for rodlike particles: vortices spontaneously form around the intruders, most pronounced at an 
optimal chiral angle where the mean curvature of particle trajectories matches the intruder 
boundary curvature, maximizing the effective force. We map the attractive and repulsive force 
regimes across chirality, propulsion, and intruder separation, offering new insights and 
principles for designing and controlling self-assembled active systems.
\end{abstract}

\maketitle

\noindent\Red{\bf{Introduction}} 
\smallskip\smallskip

\noindent Formation of patterns and structures is among the most remarkable features 
of active matter systems \cite{Cates15,Vicsek12,Shaebani20,Elgeti15}. Over the 
past two decades, significant progress has been made in understanding phenomena 
such as segregation, self-assembly, and clustering in active-passive mixtures 
\cite{Stenhammar15,SchwarzLinek12,Gokhale22}, motility-induced phase separation 
\cite{Cates15}, and the emergence of effective interactions in confined active 
fluids \cite{Paul22,Dor22} and actomyosin gels \cite{Rupprecht18,Makhija16}. 
Most studies have focused on linear (i.e., non-chiral) active systems, where 
rotational symmetry is preserved as particles reorient primarily through 
collisions or noise. In this class of active matter, the physics is governed 
by the interplay of propulsion, fluctuations, and nonlinearity and multiscale 
nature of interparticle interactions.

Nevertheless, lack of mirror symmetry is ubiquitous in both living and synthetic 
active systems. Examples include the clockwise \cite{DiLuzio05,Lauga06,Li08,
PerezIpina19} or counterclockwise \cite{DiLeonardo11} circular motion of 
flagellated bacteria near surfaces, helical swimming of marine zooplankton 
\cite{Jekely08} and sperm cells \cite{Su12,Friedrich07}, and chiral trajectories 
of synthetic self-propelled objects with asymmetric shape \cite{Kummel13,tenHagen14} 
or mass distribution \cite{Arora21} with respect to their propulsion axis. The 
coupling of translational and rotational modes of active motion is known to optimize 
navigation \cite{Shaebani20b,Shaebani22,Najafi18} and induce gearlike rolling 
in dense confined geometries \cite{Yang21,Aragones16}, suggesting that the 
coupling may also affect fluctuation-induced forces generic to nonequilibrium 
systems \cite{Aminov15,Lee17}. To get insight into how chiral active matter 
organizes into dynamic patterns and structures \cite{Liebchen17,Liebchen22}, 
a detailed understanding of the interplay between active rotation and other 
key aspects of the problem is crucial, which yet remains elusive. 

Fluctuation-induced (FI) forces act across different length scales in nonequilibrium 
systems of driven passive \cite{Shaebani12,Cattuto06,Shaebani13} or active 
\cite{Ni15,Leite16,Ray14,Feng21,Angelani11,Liu20,Harder14,Baek18,ParraRojas14} 
particles. While FI forces are typically weak in passive systems, they can become 
significantly enhanced with increasing activity \cite{Ni15}. The force strength 
depends sensitively on bath properties such as density and noise strength, as 
well as on the size, shape, orientation, and separation of intruder particles 
\cite{Shaebani12,Shaebani13,Leite16,Baek18,Ni15,Ray14,Harder14}. Moreover, 
switching between attractive and repulsive forces have been observed upon 
varying the bath density or separation and size of the intruders \cite{Shaebani12,
Ray14,Leite16,Ni15,Feng21}. Whether and how the interplay of FI forces and 
other interactions can prevent or promote formation of structures in chiral 
active matter systems has remained unexplored to date. The subject is of 
fundamental and practical importance as, for example, in biofilm formation 
and design of micro-devices \cite{DiLeonardo10} and active chiral crystals 
\cite{vanZuiden16,Tan22,Petroff15}. A major step forward is to clarify 
how chirality influences fluctuation-induced interactions, a question only 
recently touched upon \cite{Feng21}.

\begin{figure*}[t]
\centering
\includegraphics[width=0.8\linewidth]{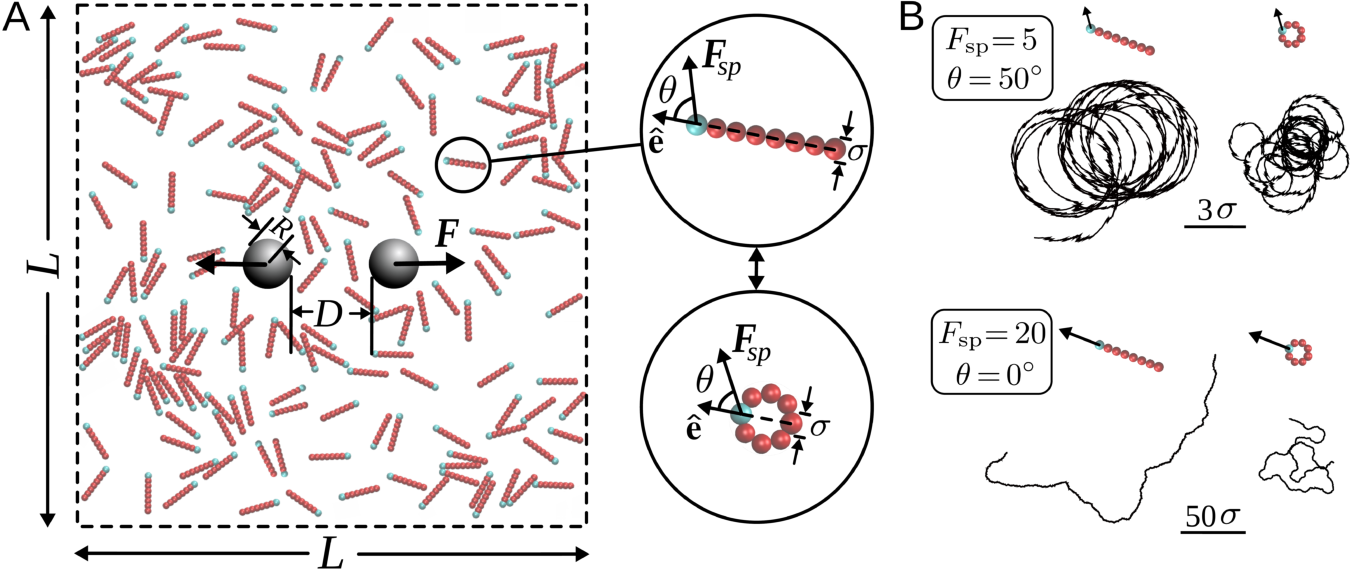}
\caption{{\bf Schematic of the simulation box.} (A) Typical snapshot of the system 
with area fraction $\phi\,{=}\,0.1$. Immobile (gray) intruders are immersed in 
an active bath consisting of either rodlike or circular rigid composites of beads 
as displayed in the insets. The self-propulsion (active) force $\fspVec$ is exerted 
on the head (blue) bead with the self-propulsion angle (chirality) $\theta$ with 
respect to the intrinsic orientation $\Vint$ of the active object. (B) Sample 
trajectories of active objects during the same time interval for different 
choices of chirality $\theta$ and magnitude of active force $\fsp$ 
(in $\funit$ units).}
\label{Fig1}
\end{figure*}
  
Here, we address this open problem by investigating the effective interactions 
between immobile intruders immersed in chiral active fluids (Fig.\,\ref{Fig1}A) 
using extensive Langevin dynamics simulations. We explore how the combination 
of chirality and elongation shapes the effective long-range FI forces beyond 
the depletion interaction range. Specifically, we demonstrate that particle 
elongation plays a pivotal role: in dense environments, pure rotation leads 
to localized spinning for circular particles, while promoting correlated 
rotations for rods. To disentangle the effects of shape and chirality, we 
compare two limits--- circular and rodlike active particles--- and show that 
rods can generate significantly stronger FI forces under identical conditions. 
Remarkably, we uncover a nonmonotonic dependence of the FI force on chirality 
in rodlike active baths, revealing an optimal chiral angle that maximizes 
interaction strength and range. We trace this optimum to a spontaneous 
collective behavior in rodlike active baths: vortex formation and disruption 
around the intruders. The interplay between the mean curvature of active 
particle trajectories and the boundary curvature of the intruders governs 
the strength of this phenomenon. When these curvatures match, the strongest 
and most persistent vortices and thus the highest particle-intruder collision 
rates occur, which enhance the FI interaction. We further establish a phase 
diagram mapping attractive and repulsive force regimes as functions of chirality, 
propulsion strength, and intruder separation. 

\smallskip\smallskip\smallskip
\noindent\Red{\bf{Chiral active fluids}} 
\smallskip\smallskip\smallskip

\noindent We construct rigid composites of touching beads of diameter $\sigma$ 
arranged either on a straight line or on a circular ring; see Fig.\,\ref{Fig1}A. 
These composite objects constitute the elements of the bath. Here, we present 
results for eight-bead composites. We consider a two dimensional system 
motivated by chiral in-plane motion of biological and synthetic agents above 
surfaces \cite{DiLuzio05,Lauga06,Li08,PerezIpina19,DiLeonardo11,Kummel13,
tenHagen14}. Periodic boundary conditions are imposed in both directions and 
two immobile circular intruders of radius $R\,{=}\,5\,\sigma$ are immersed with 
separation $D$. By exerting a self-propulsion force $\fspVec$ on each composite 
particle, the bath changes from a passive to an active fluid with a tunable 
forward propulsion. $\fspVec$ is applied on the head bead of the rod or a 
chosen bead of the circular object (blue beads in Fig.\,\ref{Fig1}A). The 
chirality is tuned by changing the angle $\theta$ between $\fspVec$ and the 
intrinsic orientation $\Vint$ of the active object (i.e.\ the line symmetrically 
dividing the composite through the blue bead). The choice of $\theta\,{=}\,0$ 
($\theta\,{\neq}\,0$) corresponds to a non-chiral (chiral) active fluid. 

The intruders and the constituent elements of the active/passive bath are 
assumed to be rigid; thus, an exclusion interaction between particles is 
introduced through the Weeks-Chandler-Andersen potential \cite{Weeks71}
\begin{equation}
U_{_\text{WCA}}(d) = \left\lbrace
\begin{array}{l l}
\!\!\!U_{_\text{LJ}}(d{-}\Delta) - U_{_\text{LJ}}(d_{\text{c}} ) 
  & \text{if $d\,{-}\,\Delta\,{<}\,d_{\text{c}}$},\\
\!\!\!0 & \text{if $d\,{-}\,\Delta\,{\geq}\,d_{\text{c}}$},
\end{array}
\right. 
\label{Eq:WCA}
\end{equation}
with $d$ being the center-to-center distance between the interacting objects, 
$U_{\text{LJ}}$ the Lennard-Johnes potential, and $d_{\text{c}}\,{=}\,2^{1/6}
\sigma$ the cut-off distance. $\Delta$ equals $0$ or $R\,{-}\,\sigma{/}2$ 
for bead-bead (of different particles) or bead-intruder interaction, respectively. 
The position $\text{\bf r}_{_i}$ of the $i$th bead of each composite object 
evolves according to the Langevin equation
\begin{equation}
{M}\,\ddot{\text{\bf r}}_{_i} = -\eta\,\dot{\text{\bf r}}_{_i} - 
{\bm\nabla}U_{_i} + {\bm\xi}_{_i} + {\bm F}\!_{_{\text{sp}{,}i}}
\,\delta_{_{h i}},
\label{Eq:LD}
\end{equation}
where $M$, $\eta$ and $U_{_i}$ are the mass of each bead, solvent friction, 
and sum of all interactions acting on the $i$th bead, respectively. The 
random force ${\bm\xi}_{_i}$ is a Gaussian white noise with zero mean and 
correlation $\langle{\xi}_{_{i,a}}(t)\,{\xi}_{_{j,b}}(t')\rangle\,{=}\,4\,
\eta\,\kbt\,\delta_{ab}\,\delta (t\,{-}\,t')$, in which $a$ and $b$ denote 
Cartesian components of the vectors, $k_{\textrm{B}}$ the Boltzmann constant, 
and $T$ the temperature. The constraint that the active force only acts 
on the head ($h$) bead is enforced by the Kronecker delta in the last 
term. A few examples of the resulting particle trajectories at area fraction 
$\phi\,{=}\,0.1$ are shown in Fig.\,\ref{Fig1}B. It is known that increasing 
self-propulsion enhances forward motion and the asymptotic diffusion 
coefficient of non-interacting active agents \cite{Nossal74,Shaebani19}, 
while inducing asymmetry in propulsion reduces the diffusion coefficient 
and leads to spiral trajectories \cite{Shaebani22b}. Similar trends are 
observed here upon varying the active force and chirality at low densities 
of the active bath. In the nonequilibrium steady state we measure the net 
force exerted by the active fluid on each intruder along the line which 
connects their centers. The force acting on the right intruder is denoted 
by ${\bm F}$ in the following. Because of large fluctuations, the force 
is measured for $10^4$ successive time intervals in the steady state and 
ensemble averaged over $10^2$ uncorrelated trajectories; see {\it Materials 
and Methods} for more details.

\begin{figure*}[t]
\centering
\includegraphics[width=0.93\linewidth]{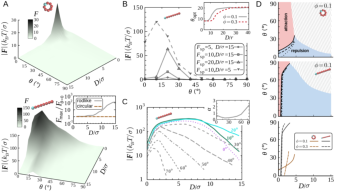}
\caption{{\bf FI forces in chiral ($\bm{\theta\,{\neq}\,0}$) active fluids.} (A) 
FI force $F$ in terms of scaled gap size $D{/}\sigma$ and chiral angle $\theta$ 
at $\fsp{=}\,10\,\funit$ and $\phi\,{=}\,0.1$ for circular (top) and rodlike (bottom) 
active objects. The inset shows the ratio between the maximum FI force $\fmax$ and 
the non-chiral FI force $\fnc$ at different values of $D{/}\sigma$ for circular and 
rodlike objects. (B) $F$ versus $\theta$ in active baths of rodlike objects with 
$\phi\,{=}\,0.1$ for different values of $\fsp$ and $D{/}\sigma$. Inset: Optimal 
chiral angle $\topt$ versus $D{/}\sigma$ for different bath densities. (C) $F$ versus 
$D$ for rodlike active particles at different values of chirality $\theta$. Other 
parameters: $\fsp{=}\,10\,\funit$, $\phi\,{=}\,0.1$. Inset: Decay exponent $\alpha$ 
of the exponential fit to the $F$-$D$ tail behavior in terms of $\theta$. (D) Phase 
diagram of force sign switching in the ($D$,$\theta$) plane for $\fsp{=}20$. The upper 
and middle panels represent the diagram for active chiral baths of circular or rodlike 
particles, respectively. The solid line marks the interface for $\fsp{=}20\,\funit$. 
The interface position is also shown for $\fsp{=}10$ (dashed line) and $5\,\funit$ 
(dotted line). The hatched zones correspond to weak FI forces where random sign 
reversals occur within the accuracy of our measurements. The lower panel represents 
a comparison between the interface positions at bath densities $\phi\,{=}\,0.1$ and 
$0.3$ for $\fsp{=}20\,\funit$.}
\label{Fig2}
\end{figure*}

\smallskip\smallskip
\noindent\Red{\bf{Long-range fluctuation-induced interactions}} 
\smallskip\smallskip

\noindent To see how the FI force is influenced by chirality, we vary the 
chiral angle $\theta$ and measure $F$ for different gap sizes $D$ between the 
intruders. Interestingly, the results reveal that the impact of $\theta$ on 
$F$ depends on the elongation of active particles. For circular particles, 
increasing $\theta$ monotonically weakens $F$ (Fig.\,\ref{Fig2}A). It means 
that for any given separation $D$, the intruders in an active fluid with 
circular constituent elements experience the maximum FI force in the non-chiral 
case ($\theta\,{=}\,0$). Contrarily, in case of rodlike active particles $F$ 
behaves nonmonotonically upon increasing $\theta$; it develops a peak at a 
nonzero chiral angle $\topt$ and decays at larger chiral angles. The inset 
of Fig.\,\ref{Fig2}A shows that the relative strengthening of $F$ compared 
to the non-chiral case is enhanced with increasing $D$. At large separations, 
where the force in the non-chiral case is weak, adopting the optimal chiral 
angle can strengthen the force by several orders of magnitude.

In Fig.\,\ref{Fig2}\;\!B, we plot $F$ versus $\theta$ for different values 
of $\fsp$ and $D$ at bath density $\phi\,{=}\,0.1$. It can be seen that the 
optimal chiral angle $\topt$ is independent of the choice of self-propulsion 
force. In addition, $\topt$ slightly grows with $D$ at small separations but 
eventually saturates to $\topt{\simeq}\,20^\circ$ at larger $D$ values (inset 
of Fig.\,\ref{Fig2}\;\!B). Similar conclusions are drawn for other values of 
bath density, though, the transition to asymptotic $\topt$ shifts to larger 
$D$ values with increasing $\phi$. 
 
The optimal chirality also remarkably enhances the range of FI interactions. 
Figure\,\ref{Fig2}\;\!C shows $F$ in terms of $D$ for different values of 
$\theta$. The tail of the FI force versus $D$ can be approximated by an 
exponential decay $F_\text{tail}\,{\sim}\,\text{e}^{-\alpha(\theta)\,D{/}\sigma}$ 
for each choice of the chiral angle. The decay slope $\alpha$ depends on 
the chirality: As shown in the inset of Fig.\,\ref{Fig2}\;\!C, by adopting 
a chiral angle around the optimal angle $\topt$, $\alpha$ reaches a minimum 
value corresponding to the longest range for the FI interaction. Moreover, 
Fig.\,\ref{Fig2}\;\!C shows a novel feature in active fluids: The FI force behaves 
nonmonotonically as a function of the separation between the intruders. It first 
develops a peak--- at a value of $D$ which can be beyond the short depletion 
range--- and then decays. This feature appears regardless of active particle 
shape, though the peak is more pronounced for elongated particles.
 
Our results also reveal that in chiral active fluids, the sign of the FI force is 
determined in general by the interplay of chirality, self-propulsion, bath density, 
and separation between the intruders. In Fig.\,\ref{Fig2}\;\!D, we present a cut 
through the phase diagram of FI force sign reversal, where the separation and 
chirality are changed. For both circular and rodlike active particles, increasing 
$\theta$ shifts the transition from attraction to repulsion to larger separations. 
$\fsp$ acts in the opposite direction, i.e., pushes the repulsive force zone to 
smaller $D$ values (more pronounced for circular active particles); see top and 
middle panels of Fig.\,\ref{Fig2}\;\!D. The impact of bath density on the force 
sign depends on the chirality of the active particles: At small values of $\theta$, 
increasing $\phi$ enhances the attractive force domain. The effect is more pronounced 
for circular active particles. In contrast, increasing $\phi$ at large chiral angles 
reduces the attraction domain to smaller separations (lower panel). We note that 
the hatched zones in the phase diagrams of Fig.\,\ref{Fig2}\;\!D mark the regions 
where the FI force is too weak ($F\,{\lesssim}\,0.1\,\funit$) and random sign 
switchings occur within the accuracy of our time-consuming measurements. Increasing 
$\fsp$ or $\phi$ strengthens the FI force and pushes the hatched zone to the right, 
i.e., larger separations. 

\smallskip\smallskip
\noindent\Red{\bf{Collision statistics around intruders}} 
\smallskip\smallskip

\noindent To uncover the physical mechanisms driving fluctuation-induced (FI) forces 
in chiral active systems, it is essential to analyze the distribution of thermodynamic 
fields around the intruders. In randomly driven passive baths, it has been shown that 
large intruder objects alter pressure field fluctuations in the confined region between 
them due to the boundary conditions they impose, resulting in effective long-range 
interactions \cite{Shaebani12,Cattuto06}. More recently, it was demonstrated that 
a probe placed near lateral walls in an active fluid experiences an effective force 
linked to the asymmetric distribution of active particles around it \cite{Paul22}. 

\begin{figure}[t]
\centering
\includegraphics[width=0.99\linewidth]{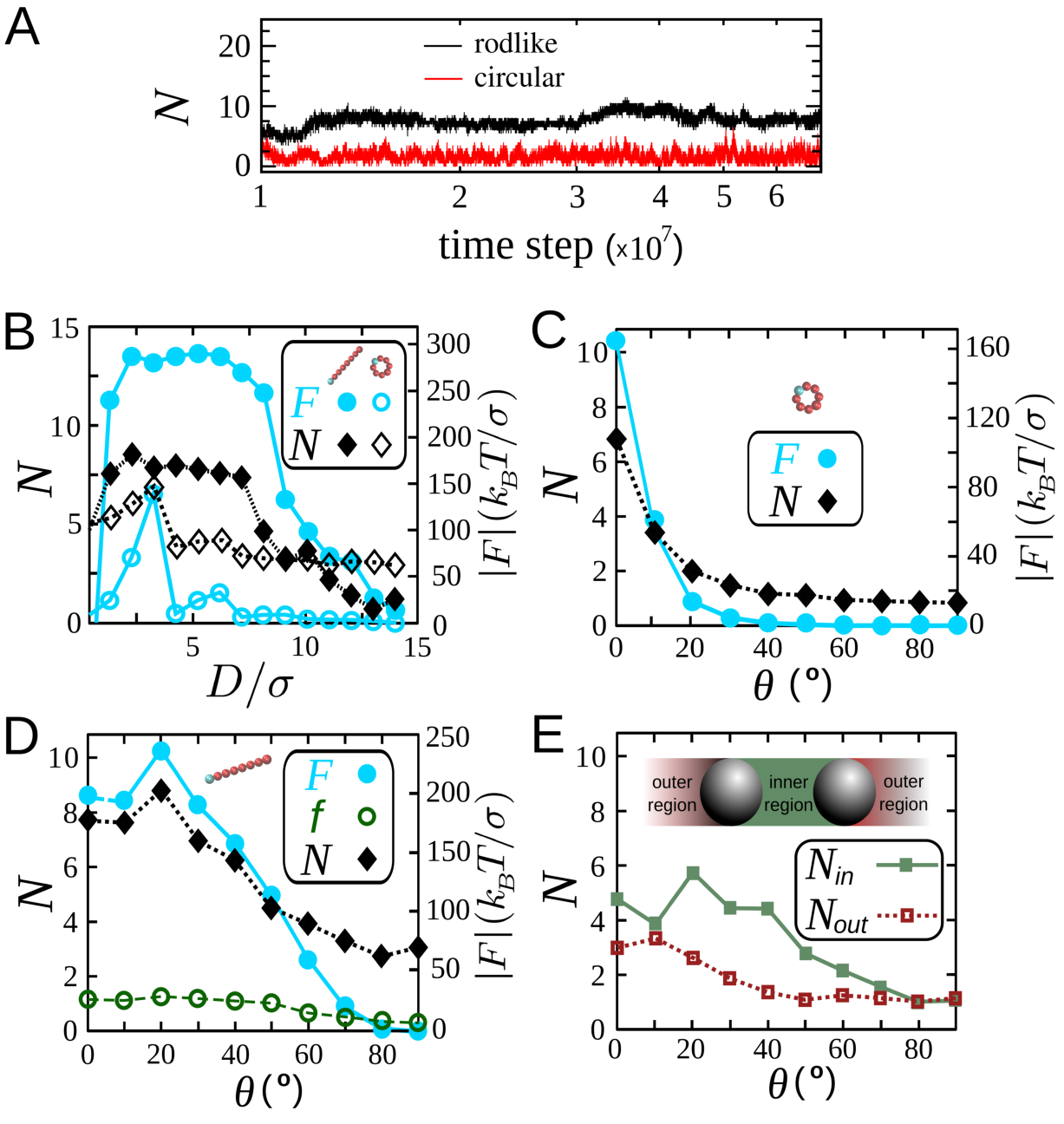}
\caption{{\bf Statistics of collisions between active particles and intruders.} (A) 
Time evolution of the number of collisions (per $10^3$ time steps) between the right 
intruder and active particles in the steady state (time steps $\!{>}10^7$), for $\phi\,
{=}\,0.3$, $D{/}\sigma\,{=}\,8$, $\fsp{=}\,20\,\funit$, and $\theta\,{=}\,20^{\circ}$. 
(B) Mean number of collisions $N$ with the right intruder and corresponding FI force 
$F$ as functions of separation $D$ for both rodlike and circular particles at $\fsp
{=}\,20\,\funit$, $\phi\,{=}\,0.1$, and $\theta\,{=}\,0$. (C,D) $N$ and $F$ versus 
chiral angle $\theta$ for $\fsp{=}\,20\,\funit$ in systems with circular particles 
($\phi\,{=}\,0.3$, $D{/}\sigma\,{=}\,3$) and rodlike particles ($\phi\,{=}\,0.1$, 
$D{/}\sigma\,{=}\,2$), respectively. Panel (D) also shows the mean magnitude of force 
per collision, $f$, in terms of $\theta$. (E) Collision counts for rodlike particles 
in panel (D), separately shown for the inner and outer regions between intruders, 
as indicated in the inset.}
\label{Fig3}
\end{figure}

We quantify the collision statistics between active particles and intruders, which 
serve as a direct measure of the asymmetry in thermodynamic fields and thus provide 
insights into the origins of FI forces. We hypothesize that the observed nonmonotonic 
dependence of the FI force on chirality or intruder separation stems from a similarly 
nonmonotonic dependence of the collision imbalance around the intruders. To test this 
hypothesis, we measure the steady-state number of particle-intruder collisions 
(Fig.\,\ref{Fig3}\;\!A). Representative examples of the dependence of the force $F$ 
and mean number of collisions $N$ on separation $D$ for both particle shapes are 
shown in Fig.\,\ref{Fig3}\;\!B, revealing a strong correlation between $F$ and $N$. 
Likewise, the nonmonotonic variation of $F$ with chirality $\theta$ in rodlike active 
fluids mirrors that of $N$. A comparison of circular and rodlike particle systems 
(Figs.\,\ref{Fig3}\;\!C,D) shows that while $N$ decays monotonically with $\theta$ 
for circular particles, it exhibits a pronounced peak for rods--- paralleling the 
behavior of $F$. This correlation establishes a direct link between $F$ and $N$. 
Furthermore, by separating collisions in the inner and outer regions in Fig.\,\ref{Fig3}\;\!E, 
we confirm that the nonmonotonic behavior of $N$ with $\theta$ arises from enhanced 
collisions in the inner region. We note that, in addition to the number of collisions, 
the mean force per collision also exhibits a slight peak at intermediate chiral angles 
(Fig.\,\ref{Fig3}\;\!D).

\begin{figure*}
\centering
\includegraphics[width=0.99\linewidth]{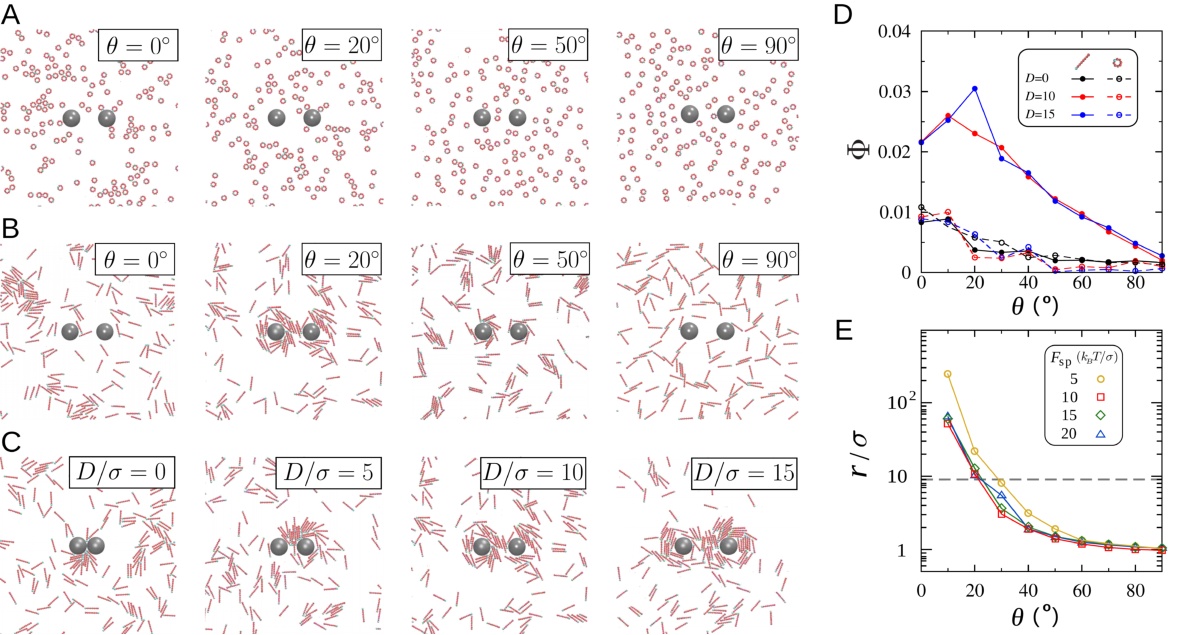}
\caption{{\bf Vortex formation around intruders in chiral active baths of rodlike 
particles.} (A-C) Typical steady-state configurations of chiral active particles 
around the intruders. The active bath has density $\phi\,{=}\,0.1$, and particles 
are driven by an active force $\fsp{=}\,10\,\funit$. In panels (A) and (B), the 
intruder separation is $D{/}\sigma\,{=}\,10$, and circular or rodlike particles 
have chiral angles $\theta\,{=}\,0$, $20$, $50$, or $90^{\circ}$. In panel (C), 
rodlike particles have $\theta\,{=}\,20^{\circ}$, and the intruder separation 
varies as $D{/}\sigma\,{=}\,0$, $5$, $10$, or $15$. (D) Vortex order parameter 
$\Phi$ as a function of chiral angle $\theta$ for circular and rodlike active 
particles at $\phi\,{=}\,0.1$, $\fsp{=}\,10\,\funit$, and different intruder 
separations. (E) Radius of curvature of the spiral trajectories (traced by the 
center of mass) of rodlike particles as a function of chiral angle, shown for 
different values of $\fsp$. The horizontal dashed line indicates the effective 
vortex radius near the intruders obtained as the intruder radius plus half the 
rod length.}
\label{Fig4}
\end{figure*}

\smallskip\smallskip
\noindent\Red{\bf{Vortex formation around intruders}} 
\smallskip\smallskip

\noindent The emergence of a peak in particle-intruder collisions at intermediate 
chiral angles $\theta\,{\approx}\,20^{\circ}$ for elongated active particles raises 
the question of its underlying mechanism. To investigate this, we examine the 
dynamics of chiral active particles in the presence of intruders. As shown in 
Fig.\,\ref{Fig4}A (see also Suppl.\,Movie\,S1), increasing $\theta$ for circular 
particles leads to a gradual transition from rotating flocks to localized spinners, 
largely independent of intruder presence. In contrast, for rodlike particles, a 
striking collective behavior arises: rods spontaneously organize into vortices 
around the intruders (Fig.\,\ref{Fig4}B and Suppl.\,Movie\,S2). We quantify this 
effect by computing the local mean velocity field ${\bm v}({\bm r})$ around the 
intruders and define the vortex order parameter as $\Phi = \frac{\langle |{\bm v} 
\cdot {\bm t}| \rangle / \langle |{\bm v}| \rangle - 2/\pi}{1 - 2/\pi}$, where 
${\bm t}$ is the unit vector tangent to the (clockwise) circular path around each 
intruder \cite{Wioland13}. $\Phi \,{=}\, 1$ corresponds to steady azimuthal circulation, 
while small values of $\Phi$ indicate a disordered chaotic flow. Vortex formation is highly 
dynamic: vortices around both intruders interfere with one another, leading to continual 
cycles of formation and disruption. This stochastic crowding around the intruders enhances 
particle-intruder collisions. Notably, $\Phi$ peaks around $\theta\,{\approx}\,20^{\circ}$ 
for rodlike active particles, coinciding with the highest collision rate and the strongest 
FI force (Fig.\,\ref{Fig4}D). For circular particles, by contrast, $\Phi$ remains weak 
and decreases with increasing $\theta$.

The dependence of vortex formation on particle elongation arises from differences in 
the dynamics of individual circular and rodlike active particles. When the propulsion 
direction and intrinsic orientation of an active particle are misaligned, two key 
effects arise. First, forward propulsion is reduced, which shortens the directional 
correlation length \cite{Sadjadi21}; this effect impacts both circular and rodlike 
particles similarly. Second, the misalignment induces a torque that leads to self-rotation 
and spiral trajectories, as illustrated in Fig.\,\ref{Fig1}B. Such torque can arise 
from viscous interactions, for example, between the cell body and flagella of swimming 
bacteria near surfaces, which alter local pressure and flow fields. The curvature of 
these trajectories depends on shape: for bacteria, the radius increases with cell-body 
length \cite{Lauga06,DiLeonardo11}, and in asymmetric artificial microswimmers, the 
curvature is largely independent of propulsion strength but governed by shape 
asymmetry \cite{Kummel13}. The mean radius of the particle trajectory $r$ can be 
roughly estimated by assuming that the translational and angular velocities are 
proportional to the forward propulsion force and the torque about the center of 
mass, respectively \cite{Lei19}. Neglecting noise, this yields $r\,{\propto}\cot
\theta$, indicating that the spiral radius is a decreasing function of $\theta$ 
and is independent of the propulsion force. By extracting the mean radius of spiral 
trajectories from simulations, traced by the center of mass, we confirm that rodlike 
particles exhibit larger spiral radii than circular ones at the same chiral angle, 
leading to more correlated dynamics and stronger FI forces. As chirality increases 
toward $\theta\,{=}\,90^{\circ}$, where propulsion vanishes, rodlike particles continue 
to exhibit collective, entangled motion due to gearlike rolling. In contrast, circular 
particles gradually transition from vortex-like patterns and rotating flocks to isolated 
spinners with no net displacement \cite{Zhang22,vanZuiden16}. As shown in Fig.\,\ref{Fig4}E 
for rodlike particles, the trajectory radius decreases with increasing chirality. Crucially, 
around $\theta\,{\approx}\,20^{\circ}$, the trajectory radius closely matches the effective 
vortex radius near the intruders--- defined as the intruder radius plus half the rod length. 
This geometric matching enhances vortex stability, maximizing particle-intruder collisions 
and thus the FI force. It can be also seen that the $\theta$-dependence of the trajectory 
radius is largely unaffected by propulsion strength, which explains why the optimal 
chiral angle $\topt$ remains largely independent of $\fsp$.

\smallskip\smallskip
\noindent\Red{\bf{Discussion and conclusion}} 
\smallskip\smallskip

\noindent Having established that vortex formation and disruption underlie the enhanced 
particle-intruder collisions and the chirality-dependent optimization of FI forces, we 
now address the origin of the observed force optimization with respect to intruder 
separation. It is known that the mean time for an active searcher to encounter a 
target--- or, equivalently, the cover time to scan a confined space--- reaches a 
minimum at an optimal intermediate confinement size relative to the persistence 
length of the searcher \cite{Shaebani20b}. Analogously, for a given propulsion 
strength $\fsp$ in a non-chiral active bath, we expect an optimal intruder separation 
$D$ that minimizes the mean hitting time of intruders (targets) by active bath 
particles (searchers) in the inner confined region between the intruders (see inset 
of Fig.\,\ref{Fig3}\;\!E). This maximizes the collision frequency and thus the FI 
force. In contrast, changes in $D$, at given $\fsp$ and $\theta$, have a smaller 
impact on collision rates in the outer region, where confinement varies less significantly.   

In chiral baths of rodlike particles, the dependence of collision statistics on intruder 
separation $D$ is strongly influenced by the ability of the system to form and sustain 
vortices. As shown in Fig.\,\ref{Fig4}\;\!C for chirality $\theta\,{=}\,20^{\circ}$ (see 
also Suppl.\,Movie\,S3), vortices cannot develop at very small separations. A minimum 
separation--- at least comparable to the length of the rodlike particles--- is required 
to allow vortex formation around the intruders and enable their mutual interaction. This 
is also evident in Fig.\,\ref{Fig4}\;\!D, where the vortex order parameter remains weak 
for small $D$, and the optimal chiral angle for rods emerges only when the separation is 
sufficiently large. This observation further explains why the optimal chirality in the 
inset of Fig.\,\ref{Fig2}\;\!B reaches a plateau only at larger values of $D$.

\begin{figure}[t]
\centering
\includegraphics[width=0.99\linewidth]{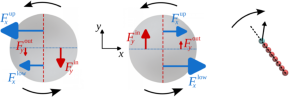}
\caption{{\bf Schematic of the FI force components.} The inner and outer (upper 
and lower) halves of each intruder are separated by the red dashed (blue dotted) 
lines. Examples of net force imbalances between different halves are: 
$\Delta F\!_x{=}|F\!\!_x^{\,\;\text{up}}\!{-}F\!\!_x^{\,\;\text{low}}|\,{\simeq}
\,40.5$ vs $\Delta F\!_y{=}|F\!\!_y^{\,\;\text{in}}\!{-}F\!\!_y^{\,\;\text{out}}|
\,{\simeq}\,27.7\,\funit$ for $\phi\,{=}\,0.1$, and $\Delta F\!_x\,{\simeq}\,21.95$ 
vs $\Delta F\!_y\,{\simeq}\,17.87\,\funit$ for $\phi\,{=}\,0.3$. Other parameters: 
$D{/}\sigma{=}3$, $\theta{=}30^{\circ}$, and $\fsp{=}20\funit$. In these examples, 
the horizontal forces are repulsive, and the perpendicular force acting on the 
right (left) intruder is upward (downward).}
\label{Fig5}
\end{figure}

Finally, we note that the unidirectional rotation of chiral active particles exerts 
a net torque on each intruder \cite{Torrik21}, inducing a gearlike rotation--- even 
in the case of a single intruder subject to isotropic collisions. We find a systematic 
imbalance in the horizontal force (along the $x$-axis in Fig.\,\ref{Fig5}) exerted on 
the upper versus lower halves of each intruder, quantified by $\Delta F\!_x{=}|F\!\!_x^{
\,\;\text{up}}\!{-}F\!\!_x^{\,\;\text{low}}|\,{>}\,0$. Likewise, there is a net vertical 
force imbalance between the inner and outer halves of each intruder, $\Delta F\!_y{=}|F\!
\!_y^{\,\;\text{in}}\!{-}F\!\!_y^{\,\;\text{out}}|\,{>}\,0$. The resulting torques from 
$\Delta F\!_x$ and $\Delta F\!_y$ act in opposite directions but do not cancel, as 
$\Delta F\!_x\,{>}\,\Delta F\!_y$. Consequently, the intruders rotate in the direction 
opposite to that of the active particles, as illustrated in Fig.\,\ref{Fig5}.

In summary, we have demonstrated that the interplay between propulsion, chirality, and 
particle shape dramatically modulates long-range fluctuation-induced forces between 
objects immersed in chiral active fluids. Specifically, we show that in systems of 
elongated active particles, the magnitude and range of the FI force can be significantly 
enhanced by tuning chirality, with an optimal chiral angle promoting vortex formation 
and maximizing interaction strength. Moreover, confinement between intruders introduces 
an optimal separation that further amplifies these interactions by increasing 
particle-intruder encounter rates. These effective forces can play a critical role 
in driving self-organization and assembly in active matter, particularly at microscales. 
Analogous phenomena have been observed in biological systems, where stress fluctuations 
in actomyosin gels induce attraction toward confining boundaries \cite{Rupprecht18}, and 
cytoskeletal dynamics drive nuclear fluctuations \cite{Makhija16}.

Our findings position chirality--- often overlooked--- as a powerful control parameter 
for engineering active systems. The coupling of rotational motion, propulsion, and geometry 
opens up new strategies for directed assembly, pattern formation, and mechanical actuation 
in systems with active chiral constituents, such as biofilms, synthetic active crystals, 
or micromachines. Looking ahead, curvature-driven mechanisms uncovered here may guide 
future studies on complex geometries, heterogeneous active mixtures, and torque-controlled 
collective dynamics in microrobotics or soft active materials.

\smallskip\smallskip
\noindent\Red{\bf{Materials and methods}} 
\smallskip\smallskip

\noindent{\bf{Simulation Method.}} Each active composite is composed of 8 
contacting spherical beads with diameter $\sigma$. The rodlike composites 
have a length of $8\,\sigma$ and the circular ones have a radius of ${\sim}\,
1.3\,\sigma$. The unequal covered area by the two types of composites requires 
that a different numbers of them be used to achieve a given occupied area fraction. 
The size of the simulation box is $L\,{=}100\,\sigma$. The Lennard-Jones potential 
in Eq.\,(\ref{Eq:WCA}) between interacting particles with center-to-center 
distance $d$ is defined as 
\begin{equation}
U_{_\text{LJ}}(d) =  4\epsilon\!\!\
\bigg[\bigg(\frac{\sigma}{d}\bigg)^{\!\!12}\!{-}\,\bigg(\frac{\sigma}{d} 
\bigg)^{\!\!6}\,\bigg],
\label{Eq:LJ}
\end{equation}
where $\epsilon$ is the potential well depth. The mass of each bead $M$, its diameter 
$\sigma$, and the potential depth $\epsilon$ are chosen as the units of mass, length, 
and energy, respectively. The solvent friction is $\eta\,{=}\,10$ in all simulations. 
The Langevin dynamics simulations are carried using the LAMMPS code \cite{LAMMPS,
Plimpton95}. The integration time step is taken to be $10^{-4}\,\tau$, with $\tau\,
{=}\,\sqrt{M \sigma^2{/}\epsilon}$ being the simulation time unit. Before starting 
to measure the FI forces, we wait for $10^{7}$ time steps to reach the steady state 
in the presence of active forces. Next, the mean FI force acting on each intruder 
during 10 time steps is measured. This quantity is averaged over $10^4$ successive 
time intervals. The process of measuring the time-averaged forces are repeated over 
an ensemble of $10^2$ uncorrelated trajectories for a further reduction of the 
fluctuations.

\noindent{\bf{Supplementary Movies.}} There are 3 supplementary movies S1 to S3, 
with the following captions: 
\smallskip

{\small
\noindent {\bf Suppl.\,Movie S1}: Typical steady-state dynamics of circular chiral active 
particles around two intruders separated by $D{/}\sigma\,{=}\,10$. The active bath 
has a density $\phi\,{=}\,0.1$. The particles are driven by an active force $\fsp{=}
\,10\,\funit$ and have a chiral angle $\theta\,{=}\,0$, $20$, $50$, or $90^{\circ}$. 
\smallskip

\noindent {\bf Suppl.\,Movie S2}: Typical steady-state dynamics of rodlike chiral active 
particles around two intruders separated by $D{/}\sigma\,{=}\,10$. The active bath 
has a density $\phi\,{=}\,0.1$. The particles are driven by an active force $\fsp{=}
\,10\,\funit$ and have a chiral angle $\theta\,{=}\,0$, $20$, $50$, or $90^{\circ}$.
\smallskip

\noindent {\bf Suppl.\,Movie S3}: Typical steady-state dynamics of rodlike chiral active 
particles around two intruders separated by $D\,{=}\,0$, $5$, $10$, or $15\,\sigma$. The 
active bath has a density $\phi\,{=}\,0.1$. The particles are driven by an active force 
$\fsp{=}\,10\,\funit$ and have a chiral angle $\theta\,{=}\,20^{\circ}$.
\smallskip

\smallskip\smallskip
\noindent\Red{\bf{Acknowledgements}} 
\smallskip\smallskip

\noindent We thank Hartmut L\"owen for fruitful discussions and Rudolf Podgornik for reading 
the manuscript. R.S.\ acknowledges support by the Deutsche Forschungsgemeinschaft (DFG) 
through Collaborative Research Center SFB 1027 and by the Young Investigator Grant of 
Saarland University, Grant No.\ 7410110401.

\smallskip\smallskip
\noindent\Red{\bf{Author contributions}} 
\smallskip\smallskip

\noindent J.S. and R.S. designed research; J.S., H.F., and H.K. designed and performed 
simulations; all authors contributed to analysis and interpretation of the results; 
R.S. wrote the paper; J.S. and R.S. revised the paper. Correspondence and requests 
for materials should be addressed to jalal@ipm.ir or shaebani@lusi.uni-sb.de. 

\smallskip\smallskip
\noindent\Red{\bf{Competing interests}} 
\smallskip\smallskip

\noindent The authors declare that they have no competing interests.

\smallskip\smallskip
\noindent\Red{\bf{Data and materials availability}} 
\smallskip\smallskip

\noindent All data needed to evaluate the conclusions in the paper are present in the 
paper and/or the Supplementary Materials.

\bibliography{Refs-ActiveCasimir-CirularMotion}

\begin{thebibliography}{58}%
\makeatletter
\providecommand \@ifxundefined [1]{%
 \@ifx{#1\undefined}
}%
\providecommand \@ifnum [1]{%
 \ifnum #1\expandafter \@firstoftwo
 \else \expandafter \@secondoftwo
 \fi
}%
\providecommand \@ifx [1]{%
 \ifx #1\expandafter \@firstoftwo
 \else \expandafter \@secondoftwo
 \fi
}%
\providecommand \natexlab [1]{#1}%
\providecommand \enquote  [1]{``#1''}%
\providecommand \bibnamefont  [1]{#1}%
\providecommand \bibfnamefont [1]{#1}%
\providecommand \citenamefont [1]{#1}%
\providecommand \href@noop [0]{\@secondoftwo}%
\providecommand \href [0]{\begingroup \@sanitize@url \@href}%
\providecommand \@href[1]{\@@startlink{#1}\@@href}%
\providecommand \@@href[1]{\endgroup#1\@@endlink}%
\providecommand \@sanitize@url [0]{\catcode `\\12\catcode `\$12\catcode
  `\&12\catcode `\#12\catcode `\^12\catcode `\_12\catcode `\%12\relax}%
\providecommand \@@startlink[1]{}%
\providecommand \@@endlink[0]{}%
\providecommand \url  [0]{\begingroup\@sanitize@url \@url }%
\providecommand \@url [1]{\endgroup\@href {#1}{\urlprefix }}%
\providecommand \urlprefix  [0]{URL }%
\providecommand \Eprint [0]{\href }%
\providecommand \doibase [0]{http://dx.doi.org/}%
\providecommand \selectlanguage [0]{\@gobble}%
\providecommand \bibinfo  [0]{\@secondoftwo}%
\providecommand \bibfield  [0]{\@secondoftwo}%
\providecommand \translation [1]{[#1]}%
\providecommand \BibitemOpen [0]{}%
\providecommand \bibitemStop [0]{}%
\providecommand \bibitemNoStop [0]{.\EOS\space}%
\providecommand \EOS [0]{\spacefactor3000\relax}%
\providecommand \BibitemShut  [1]{\csname bibitem#1\endcsname}%
\let\auto@bib@innerbib\@empty
\bibitem [{\citenamefont {Cates}\ and\ \citenamefont
  {Tailleur}(2015)}]{Cates15}%
  \BibitemOpen
  \bibfield  {author} {\bibinfo {author} {\bibfnamefont {Michael~E.}\
  \bibnamefont {Cates}}\ and\ \bibinfo {author} {\bibfnamefont {Julien}\
  \bibnamefont {Tailleur}},\ }\bibfield  {title} {\enquote {\bibinfo {title}
  {Motility-induced phase separation},}\ }\href@noop {} {\bibfield  {journal}
  {\bibinfo  {journal} {Annu. Rev. Condens. Matter Phys.}\ }\textbf {\bibinfo
  {volume} {6}},\ \bibinfo {pages} {219--244} (\bibinfo {year}
  {2015})}\BibitemShut {NoStop}%
\bibitem [{\citenamefont {Vicsek}\ and\ \citenamefont
  {Zafeiris}(2012)}]{Vicsek12}%
  \BibitemOpen
  \bibfield  {author} {\bibinfo {author} {\bibfnamefont {Tam\'as}\ \bibnamefont
  {Vicsek}}\ and\ \bibinfo {author} {\bibfnamefont {Anna}\ \bibnamefont
  {Zafeiris}},\ }\bibfield  {title} {\enquote {\bibinfo {title} {Collective
  motion},}\ }\href@noop {} {\bibfield  {journal} {\bibinfo  {journal} {Phys.
  Rep.}\ }\textbf {\bibinfo {volume} {517}},\ \bibinfo {pages} {71--140}
  (\bibinfo {year} {2012})}\BibitemShut {NoStop}%
\bibitem [{\citenamefont {Shaebani}\ \emph
  {et~al.}(2020{\natexlab{a}})\citenamefont {Shaebani}, \citenamefont
  {Wysocki}, \citenamefont {Winkler}, \citenamefont {Gompper},\ and\
  \citenamefont {Rieger}}]{Shaebani20}%
  \BibitemOpen
  \bibfield  {author} {\bibinfo {author} {\bibfnamefont {M.~Reza}\ \bibnamefont
  {Shaebani}}, \bibinfo {author} {\bibfnamefont {Adam}\ \bibnamefont
  {Wysocki}}, \bibinfo {author} {\bibfnamefont {Roland~G.}\ \bibnamefont
  {Winkler}}, \bibinfo {author} {\bibfnamefont {Gerhard}\ \bibnamefont
  {Gompper}}, \ and\ \bibinfo {author} {\bibfnamefont {Heiko}\ \bibnamefont
  {Rieger}},\ }\bibfield  {title} {\enquote {\bibinfo {title} {Computational
  models for active matter},}\ }\href@noop {} {\bibfield  {journal} {\bibinfo
  {journal} {Nat. Rev. Phys.}\ }\textbf {\bibinfo {volume} {2}},\ \bibinfo
  {pages} {181--199} (\bibinfo {year} {2020}{\natexlab{a}})}\BibitemShut
  {NoStop}%
\bibitem [{\citenamefont {Elgeti}\ \emph {et~al.}(2015)\citenamefont {Elgeti},
  \citenamefont {Winkler},\ and\ \citenamefont {Gompper}}]{Elgeti15}%
  \BibitemOpen
  \bibfield  {author} {\bibinfo {author} {\bibfnamefont {J}~\bibnamefont
  {Elgeti}}, \bibinfo {author} {\bibfnamefont {R~G}\ \bibnamefont {Winkler}}, \
  and\ \bibinfo {author} {\bibfnamefont {G}~\bibnamefont {Gompper}},\
  }\bibfield  {title} {\enquote {\bibinfo {title} {Physics of
  microswimmers-single particle motion and collective behavior: a review},}\
  }\href@noop {} {\bibfield  {journal} {\bibinfo  {journal} {Rep. Prog. Phys.}\
  }\textbf {\bibinfo {volume} {78}},\ \bibinfo {pages} {056601} (\bibinfo
  {year} {2015})}\BibitemShut {NoStop}%
\bibitem [{\citenamefont {{Stenhammar, J., et al.}}(2015)}]{Stenhammar15}%
  \BibitemOpen
  \bibfield  {author} {\bibinfo {author} {\bibnamefont {{Stenhammar, J., et
  al.}}},\ }\bibfield  {title} {\enquote {\bibinfo {title} {Activity-induced
  phase separation and self-assembly in mixtures of active and passive
  particles},}\ }\href@noop {} {\bibfield  {journal} {\bibinfo  {journal}
  {Phys. Rev. Lett.}\ }\textbf {\bibinfo {volume} {114}},\ \bibinfo {pages}
  {018301} (\bibinfo {year} {2015})}\BibitemShut {NoStop}%
\bibitem [{\citenamefont {Schwarz-Linek}\ \emph {et~al.}(2012)\citenamefont
  {Schwarz-Linek}, \citenamefont {Valeriani}, \citenamefont {Cacciuto},
  \citenamefont {Cates}, \citenamefont {Marenduzzo}, \citenamefont {Morozov},\
  and\ \citenamefont {Poon}}]{SchwarzLinek12}%
  \BibitemOpen
  \bibfield  {author} {\bibinfo {author} {\bibfnamefont {J.}~\bibnamefont
  {Schwarz-Linek}}, \bibinfo {author} {\bibfnamefont {C.}~\bibnamefont
  {Valeriani}}, \bibinfo {author} {\bibfnamefont {A.}~\bibnamefont {Cacciuto}},
  \bibinfo {author} {\bibfnamefont {M.~E.}\ \bibnamefont {Cates}}, \bibinfo
  {author} {\bibfnamefont {D.}~\bibnamefont {Marenduzzo}}, \bibinfo {author}
  {\bibfnamefont {A.~N.}\ \bibnamefont {Morozov}}, \ and\ \bibinfo {author}
  {\bibfnamefont {W.~C.~K.}\ \bibnamefont {Poon}},\ }\bibfield  {title}
  {\enquote {\bibinfo {title} {Phase separation and rotor self-assembly in
  active particle suspensions},}\ }\href@noop {} {\bibfield  {journal}
  {\bibinfo  {journal} {Proc. Natl. Acad. Sci. USA}\ }\textbf {\bibinfo
  {volume} {109}},\ \bibinfo {pages} {4052--4057} (\bibinfo {year}
  {2012})}\BibitemShut {NoStop}%
\bibitem [{\citenamefont {Gokhale}\ \emph {et~al.}(2022)\citenamefont
  {Gokhale}, \citenamefont {Li}, \citenamefont {Solon}, \citenamefont {Gore},\
  and\ \citenamefont {Fakhri}}]{Gokhale22}%
  \BibitemOpen
  \bibfield  {author} {\bibinfo {author} {\bibfnamefont {Shreyas}\ \bibnamefont
  {Gokhale}}, \bibinfo {author} {\bibfnamefont {Junang}\ \bibnamefont {Li}},
  \bibinfo {author} {\bibfnamefont {Alexandre}\ \bibnamefont {Solon}}, \bibinfo
  {author} {\bibfnamefont {Jeff}\ \bibnamefont {Gore}}, \ and\ \bibinfo
  {author} {\bibfnamefont {Nikta}\ \bibnamefont {Fakhri}},\ }\bibfield  {title}
  {\enquote {\bibinfo {title} {Dynamic clustering of passive colloids in dense
  suspensions of motile bacteria},}\ }\href@noop {} {\bibfield  {journal}
  {\bibinfo  {journal} {Phys. Rev. E}\ }\textbf {\bibinfo {volume} {105}},\
  \bibinfo {pages} {054605} (\bibinfo {year} {2022})}\BibitemShut {NoStop}%
\bibitem [{\citenamefont {Paul}\ \emph {et~al.}(2022)\citenamefont {Paul},
  \citenamefont {Jayaram}, \citenamefont {Narinder}, \citenamefont {Speck},\
  and\ \citenamefont {Bechinger}}]{Paul22}%
  \BibitemOpen
  \bibfield  {author} {\bibinfo {author} {\bibfnamefont {Shuvojit}\
  \bibnamefont {Paul}}, \bibinfo {author} {\bibfnamefont {Ashreya}\
  \bibnamefont {Jayaram}}, \bibinfo {author} {\bibfnamefont {N}~\bibnamefont
  {Narinder}}, \bibinfo {author} {\bibfnamefont {Thomas}\ \bibnamefont
  {Speck}}, \ and\ \bibinfo {author} {\bibfnamefont {Clemens}\ \bibnamefont
  {Bechinger}},\ }\bibfield  {title} {\enquote {\bibinfo {title} {Force
  generation in confined active fluids: The role of microstructure},}\
  }\href@noop {} {\bibfield  {journal} {\bibinfo  {journal} {Phys. Rev. Lett.}\
  }\textbf {\bibinfo {volume} {129}},\ \bibinfo {pages} {058001} (\bibinfo
  {year} {2022})}\BibitemShut {NoStop}%
\bibitem [{\citenamefont {Dor}\ \emph {et~al.}(2022)\citenamefont {Dor},
  \citenamefont {Kafri}, \citenamefont {Kardar},\ and\ \citenamefont
  {Tailleur}}]{Dor22}%
  \BibitemOpen
  \bibfield  {author} {\bibinfo {author} {\bibfnamefont {Ydan~Ben}\
  \bibnamefont {Dor}}, \bibinfo {author} {\bibfnamefont {Yariv}\ \bibnamefont
  {Kafri}}, \bibinfo {author} {\bibfnamefont {Mehran}\ \bibnamefont {Kardar}},
  \ and\ \bibinfo {author} {\bibfnamefont {Julien}\ \bibnamefont {Tailleur}},\
  }\bibfield  {title} {\enquote {\bibinfo {title} {Passive objects in confined
  active fluids: A localization transition},}\ }\href@noop {} {\bibfield
  {journal} {\bibinfo  {journal} {Phys. Rev. E}\ }\textbf {\bibinfo {volume}
  {106}},\ \bibinfo {pages} {044604} (\bibinfo {year} {2022})}\BibitemShut
  {NoStop}%
\bibitem [{\citenamefont {Rupprecht}\ \emph {et~al.}(2018)\citenamefont
  {Rupprecht}, \citenamefont {Singh~Vishen}, \citenamefont {Shivashankar},
  \citenamefont {Rao},\ and\ \citenamefont {Prost}}]{Rupprecht18}%
  \BibitemOpen
  \bibfield  {author} {\bibinfo {author} {\bibfnamefont {J.-F.}\ \bibnamefont
  {Rupprecht}}, \bibinfo {author} {\bibfnamefont {A.}~\bibnamefont
  {Singh~Vishen}}, \bibinfo {author} {\bibfnamefont {G.~V.}\ \bibnamefont
  {Shivashankar}}, \bibinfo {author} {\bibfnamefont {M.}~\bibnamefont {Rao}}, \
  and\ \bibinfo {author} {\bibfnamefont {J.}~\bibnamefont {Prost}},\ }\bibfield
   {title} {\enquote {\bibinfo {title} {Maximal fluctuations of confined
  actomyosin gels: Dynamics of the cell nucleus},}\ }\href@noop {} {\bibfield
  {journal} {\bibinfo  {journal} {Phys. Rev. Lett.}\ }\textbf {\bibinfo
  {volume} {120}},\ \bibinfo {pages} {098001} (\bibinfo {year}
  {2018})}\BibitemShut {NoStop}%
\bibitem [{\citenamefont {Makhija}\ \emph {et~al.}(2016)\citenamefont
  {Makhija}, \citenamefont {Jokhun},\ and\ \citenamefont
  {Shivashankar}}]{Makhija16}%
  \BibitemOpen
  \bibfield  {author} {\bibinfo {author} {\bibfnamefont {Ekta}\ \bibnamefont
  {Makhija}}, \bibinfo {author} {\bibfnamefont {D.~S.}\ \bibnamefont {Jokhun}},
  \ and\ \bibinfo {author} {\bibfnamefont {G.~V.}\ \bibnamefont
  {Shivashankar}},\ }\bibfield  {title} {\enquote {\bibinfo {title} {Nuclear
  deformability and telomere dynamics are regulated by cell geometric
  constraints},}\ }\href@noop {} {\bibfield  {journal} {\bibinfo  {journal}
  {Proc. Natl. Acad. Sci. USA}\ }\textbf {\bibinfo {volume} {113}},\ \bibinfo
  {pages} {E32--E40} (\bibinfo {year} {2016})}\BibitemShut {NoStop}%
\bibitem [{\citenamefont {DiLuzio}\ \emph {et~al.}(2005)\citenamefont
  {DiLuzio}, \citenamefont {Turner}, \citenamefont {Mayer}, \citenamefont
  {Garstecki}, \citenamefont {Weibel}, \citenamefont {Berg},\ and\
  \citenamefont {Whitesides}}]{DiLuzio05}%
  \BibitemOpen
  \bibfield  {author} {\bibinfo {author} {\bibfnamefont {Willow~R.}\
  \bibnamefont {DiLuzio}}, \bibinfo {author} {\bibfnamefont {Linda}\
  \bibnamefont {Turner}}, \bibinfo {author} {\bibfnamefont {Michael}\
  \bibnamefont {Mayer}}, \bibinfo {author} {\bibfnamefont {Piotr}\ \bibnamefont
  {Garstecki}}, \bibinfo {author} {\bibfnamefont {Douglas~B.}\ \bibnamefont
  {Weibel}}, \bibinfo {author} {\bibfnamefont {Howard~C.}\ \bibnamefont
  {Berg}}, \ and\ \bibinfo {author} {\bibfnamefont {George~M.}\ \bibnamefont
  {Whitesides}},\ }\bibfield  {title} {\enquote {\bibinfo {title} {Escherichia
  coli swim on the right-hand side},}\ }\href@noop {} {\bibfield  {journal}
  {\bibinfo  {journal} {Nature}\ }\textbf {\bibinfo {volume} {435}},\ \bibinfo
  {pages} {1271--1274} (\bibinfo {year} {2005})}\BibitemShut {NoStop}%
\bibitem [{\citenamefont {Lauga}\ \emph {et~al.}(2006)\citenamefont {Lauga},
  \citenamefont {DiLuzio}, \citenamefont {Whitesides},\ and\ \citenamefont
  {Stone}}]{Lauga06}%
  \BibitemOpen
  \bibfield  {author} {\bibinfo {author} {\bibfnamefont {Eric}\ \bibnamefont
  {Lauga}}, \bibinfo {author} {\bibfnamefont {Willow~R.}\ \bibnamefont
  {DiLuzio}}, \bibinfo {author} {\bibfnamefont {George~M.}\ \bibnamefont
  {Whitesides}}, \ and\ \bibinfo {author} {\bibfnamefont {Howard~A.}\
  \bibnamefont {Stone}},\ }\bibfield  {title} {\enquote {\bibinfo {title}
  {Swimming in circles: Motion of bacteria near solid boundaries},}\
  }\href@noop {} {\bibfield  {journal} {\bibinfo  {journal} {Biophys. J.}\
  }\textbf {\bibinfo {volume} {90}},\ \bibinfo {pages} {400--412} (\bibinfo
  {year} {2006})}\BibitemShut {NoStop}%
\bibitem [{\citenamefont {Li}\ \emph {et~al.}(2008)\citenamefont {Li},
  \citenamefont {Tam},\ and\ \citenamefont {Tang}}]{Li08}%
  \BibitemOpen
  \bibfield  {author} {\bibinfo {author} {\bibfnamefont {Guanglai}\
  \bibnamefont {Li}}, \bibinfo {author} {\bibfnamefont {Lick-Kong}\
  \bibnamefont {Tam}}, \ and\ \bibinfo {author} {\bibfnamefont {Jay~X.}\
  \bibnamefont {Tang}},\ }\bibfield  {title} {\enquote {\bibinfo {title}
  {Amplified effect of brownian motion in bacterial near-surface swimming},}\
  }\href@noop {} {\bibfield  {journal} {\bibinfo  {journal} {Proc. Natl. Acad.
  Sci. USA}\ }\textbf {\bibinfo {volume} {105}},\ \bibinfo {pages}
  {18355--18359} (\bibinfo {year} {2008})}\BibitemShut {NoStop}%
\bibitem [{\citenamefont {Perez~Ipina}\ \emph {et~al.}(2019)\citenamefont
  {Perez~Ipina}, \citenamefont {Otte}, \citenamefont {Pontier-Bres},
  \citenamefont {Czerucka},\ and\ \citenamefont {Peruani}}]{PerezIpina19}%
  \BibitemOpen
  \bibfield  {author} {\bibinfo {author} {\bibfnamefont {Emiliano}\
  \bibnamefont {Perez~Ipina}}, \bibinfo {author} {\bibfnamefont {Stefan}\
  \bibnamefont {Otte}}, \bibinfo {author} {\bibfnamefont {Rodolphe}\
  \bibnamefont {Pontier-Bres}}, \bibinfo {author} {\bibfnamefont {Dorota}\
  \bibnamefont {Czerucka}}, \ and\ \bibinfo {author} {\bibfnamefont {Fernando}\
  \bibnamefont {Peruani}},\ }\bibfield  {title} {\enquote {\bibinfo {title}
  {Bacteria display optimal transport near surfaces},}\ }\href@noop {}
  {\bibfield  {journal} {\bibinfo  {journal} {Nat. Phys.}\ }\textbf {\bibinfo
  {volume} {15}},\ \bibinfo {pages} {610--615} (\bibinfo {year}
  {2019})}\BibitemShut {NoStop}%
\bibitem [{\citenamefont {Di~Leonardo}\ \emph {et~al.}(2011)\citenamefont
  {Di~Leonardo}, \citenamefont {Dell'Arciprete}, \citenamefont {Angelani},\
  and\ \citenamefont {Iebba}}]{DiLeonardo11}%
  \BibitemOpen
  \bibfield  {author} {\bibinfo {author} {\bibfnamefont {R.}~\bibnamefont
  {Di~Leonardo}}, \bibinfo {author} {\bibfnamefont {D.}~\bibnamefont
  {Dell'Arciprete}}, \bibinfo {author} {\bibfnamefont {L.}~\bibnamefont
  {Angelani}}, \ and\ \bibinfo {author} {\bibfnamefont {V.}~\bibnamefont
  {Iebba}},\ }\bibfield  {title} {\enquote {\bibinfo {title} {Swimming with an
  image},}\ }\href@noop {} {\bibfield  {journal} {\bibinfo  {journal} {Phys.
  Rev. Lett.}\ }\textbf {\bibinfo {volume} {106}},\ \bibinfo {pages} {038101}
  (\bibinfo {year} {2011})}\BibitemShut {NoStop}%
\bibitem [{\citenamefont {J{\'e}kely}\ \emph {et~al.}(2008)\citenamefont
  {J{\'e}kely}, \citenamefont {Colombelli}, \citenamefont {Hausen},
  \citenamefont {Guy}, \citenamefont {Stelzer}, \citenamefont
  {N{\'e}d{\'e}lec},\ and\ \citenamefont {Arendt}}]{Jekely08}%
  \BibitemOpen
  \bibfield  {author} {\bibinfo {author} {\bibfnamefont {G{\'a}sp{\'a}r}\
  \bibnamefont {J{\'e}kely}}, \bibinfo {author} {\bibfnamefont {Julien}\
  \bibnamefont {Colombelli}}, \bibinfo {author} {\bibfnamefont {Harald}\
  \bibnamefont {Hausen}}, \bibinfo {author} {\bibfnamefont {Keren}\
  \bibnamefont {Guy}}, \bibinfo {author} {\bibfnamefont {Ernst}\ \bibnamefont
  {Stelzer}}, \bibinfo {author} {\bibfnamefont {Fran{\c{c}}ois}\ \bibnamefont
  {N{\'e}d{\'e}lec}}, \ and\ \bibinfo {author} {\bibfnamefont {Detlev}\
  \bibnamefont {Arendt}},\ }\bibfield  {title} {\enquote {\bibinfo {title}
  {Mechanism of phototaxis in marine zooplankton},}\ }\href@noop {} {\bibfield
  {journal} {\bibinfo  {journal} {Nature}\ }\textbf {\bibinfo {volume} {456}},\
  \bibinfo {pages} {395--399} (\bibinfo {year} {2008})}\BibitemShut {NoStop}%
\bibitem [{\citenamefont {Su}\ \emph {et~al.}(2012)\citenamefont {Su},
  \citenamefont {Xue},\ and\ \citenamefont {Ozcan}}]{Su12}%
  \BibitemOpen
  \bibfield  {author} {\bibinfo {author} {\bibfnamefont {Ting-Wei}\
  \bibnamefont {Su}}, \bibinfo {author} {\bibfnamefont {Liang}\ \bibnamefont
  {Xue}}, \ and\ \bibinfo {author} {\bibfnamefont {Aydogan}\ \bibnamefont
  {Ozcan}},\ }\bibfield  {title} {\enquote {\bibinfo {title} {High-throughput
  lensfree 3d tracking of human sperms reveals rare statistics of helical
  trajectories},}\ }\href@noop {} {\bibfield  {journal} {\bibinfo  {journal}
  {Proc. Natl. Acad. Sci. USA}\ }\textbf {\bibinfo {volume} {109}},\ \bibinfo
  {pages} {16018--16022} (\bibinfo {year} {2012})}\BibitemShut {NoStop}%
\bibitem [{\citenamefont {Friedrich}\ and\ \citenamefont
  {J\"ulicher}(2007)}]{Friedrich07}%
  \BibitemOpen
  \bibfield  {author} {\bibinfo {author} {\bibfnamefont {Benjamin~M.}\
  \bibnamefont {Friedrich}}\ and\ \bibinfo {author} {\bibfnamefont {Frank}\
  \bibnamefont {J\"ulicher}},\ }\bibfield  {title} {\enquote {\bibinfo {title}
  {Chemotaxis of sperm cells},}\ }\href@noop {} {\bibfield  {journal} {\bibinfo
   {journal} {Proc. Natl. Acad. Sci. USA}\ }\textbf {\bibinfo {volume} {104}},\
  \bibinfo {pages} {13256--13261} (\bibinfo {year} {2007})}\BibitemShut
  {NoStop}%
\bibitem [{\citenamefont {K\"ummel}\ \emph {et~al.}(2013)\citenamefont
  {K\"ummel}, \citenamefont {ten Hagen}, \citenamefont {Wittkowski},
  \citenamefont {Buttinoni}, \citenamefont {Eichhorn}, \citenamefont {Volpe},
  \citenamefont {L\"owen},\ and\ \citenamefont {Bechinger}}]{Kummel13}%
  \BibitemOpen
  \bibfield  {author} {\bibinfo {author} {\bibfnamefont {Felix}\ \bibnamefont
  {K\"ummel}}, \bibinfo {author} {\bibfnamefont {Borge}\ \bibnamefont {ten
  Hagen}}, \bibinfo {author} {\bibfnamefont {Raphael}\ \bibnamefont
  {Wittkowski}}, \bibinfo {author} {\bibfnamefont {Ivo}\ \bibnamefont
  {Buttinoni}}, \bibinfo {author} {\bibfnamefont {Ralf}\ \bibnamefont
  {Eichhorn}}, \bibinfo {author} {\bibfnamefont {Giovanni}\ \bibnamefont
  {Volpe}}, \bibinfo {author} {\bibfnamefont {Hartmut}\ \bibnamefont
  {L\"owen}}, \ and\ \bibinfo {author} {\bibfnamefont {Clemens}\ \bibnamefont
  {Bechinger}},\ }\bibfield  {title} {\enquote {\bibinfo {title} {Circular
  motion of asymmetric self-propelling particles},}\ }\href@noop {} {\bibfield
  {journal} {\bibinfo  {journal} {Phys. Rev. Lett.}\ }\textbf {\bibinfo
  {volume} {110}},\ \bibinfo {pages} {198302} (\bibinfo {year}
  {2013})}\BibitemShut {NoStop}%
\bibitem [{\citenamefont {ten Hagen}\ \emph {et~al.}(2014)\citenamefont {ten
  Hagen}, \citenamefont {K{\"u}mmel}, \citenamefont {Wittkowski}, \citenamefont
  {Takagi}, \citenamefont {L{\"o}wen},\ and\ \citenamefont
  {Bechinger}}]{tenHagen14}%
  \BibitemOpen
  \bibfield  {author} {\bibinfo {author} {\bibfnamefont {Borge}\ \bibnamefont
  {ten Hagen}}, \bibinfo {author} {\bibfnamefont {Felix}\ \bibnamefont
  {K{\"u}mmel}}, \bibinfo {author} {\bibfnamefont {Raphael}\ \bibnamefont
  {Wittkowski}}, \bibinfo {author} {\bibfnamefont {Daisuke}\ \bibnamefont
  {Takagi}}, \bibinfo {author} {\bibfnamefont {Hartmut}\ \bibnamefont
  {L{\"o}wen}}, \ and\ \bibinfo {author} {\bibfnamefont {Clemens}\ \bibnamefont
  {Bechinger}},\ }\bibfield  {title} {\enquote {\bibinfo {title} {Gravitaxis of
  asymmetric self-propelled colloidal particles},}\ }\href@noop {} {\bibfield
  {journal} {\bibinfo  {journal} {Nat. Commun.}\ }\textbf {\bibinfo {volume}
  {5}},\ \bibinfo {pages} {4829} (\bibinfo {year} {2014})}\BibitemShut
  {NoStop}%
\bibitem [{\citenamefont {Arora}\ \emph {et~al.}(2021)\citenamefont {Arora},
  \citenamefont {Sood},\ and\ \citenamefont {Ganapathy}}]{Arora21}%
  \BibitemOpen
  \bibfield  {author} {\bibinfo {author} {\bibfnamefont {Pragya}\ \bibnamefont
  {Arora}}, \bibinfo {author} {\bibfnamefont {A.~K.}\ \bibnamefont {Sood}}, \
  and\ \bibinfo {author} {\bibfnamefont {Rajesh}\ \bibnamefont {Ganapathy}},\
  }\bibfield  {title} {\enquote {\bibinfo {title} {Emergent stereoselective
  interactions and self-recognition in polar chiral active ellipsoids},}\
  }\href@noop {} {\bibfield  {journal} {\bibinfo  {journal} {Sci. Adv.}\
  }\textbf {\bibinfo {volume} {7}},\ \bibinfo {pages} {eabd0331} (\bibinfo
  {year} {2021})}\BibitemShut {NoStop}%
\bibitem [{\citenamefont {Shaebani}\ \emph
  {et~al.}(2020{\natexlab{b}})\citenamefont {Shaebani}, \citenamefont {Jose},
  \citenamefont {Santen}, \citenamefont {Stankevicins},\ and\ \citenamefont
  {Lautenschl\"ager}}]{Shaebani20b}%
  \BibitemOpen
  \bibfield  {author} {\bibinfo {author} {\bibfnamefont {M.~Reza}\ \bibnamefont
  {Shaebani}}, \bibinfo {author} {\bibfnamefont {Robin}\ \bibnamefont {Jose}},
  \bibinfo {author} {\bibfnamefont {Ludger}\ \bibnamefont {Santen}}, \bibinfo
  {author} {\bibfnamefont {Luiza}\ \bibnamefont {Stankevicins}}, \ and\
  \bibinfo {author} {\bibfnamefont {Franziska}\ \bibnamefont
  {Lautenschl\"ager}},\ }\bibfield  {title} {\enquote {\bibinfo {title}
  {Persistence-speed coupling enhances the search efficiency of migrating
  immune cells},}\ }\href@noop {} {\bibfield  {journal} {\bibinfo  {journal}
  {Phys. Rev. Lett.}\ }\textbf {\bibinfo {volume} {125}},\ \bibinfo {pages}
  {268102} (\bibinfo {year} {2020}{\natexlab{b}})}\BibitemShut {NoStop}%
\bibitem [{\citenamefont {Shaebani}\ \emph
  {et~al.}(2022{\natexlab{a}})\citenamefont {Shaebani}, \citenamefont {Piel},\
  and\ \citenamefont {Lautenschl\"ager}}]{Shaebani22}%
  \BibitemOpen
  \bibfield  {author} {\bibinfo {author} {\bibfnamefont {M.~Reza}\ \bibnamefont
  {Shaebani}}, \bibinfo {author} {\bibfnamefont {Matthieu}\ \bibnamefont
  {Piel}}, \ and\ \bibinfo {author} {\bibfnamefont {Franziska}\ \bibnamefont
  {Lautenschl\"ager}},\ }\bibfield  {title} {\enquote {\bibinfo {title}
  {Distinct speed and direction memories of migrating dendritic cells diversify
  their search strategies},}\ }\href@noop {} {\bibfield  {journal} {\bibinfo
  {journal} {Biophys. J.}\ }\textbf {\bibinfo {volume} {121}},\ \bibinfo
  {pages} {4099--4108} (\bibinfo {year} {2022}{\natexlab{a}})}\BibitemShut
  {NoStop}%
\bibitem [{\citenamefont {{Najafi, J., et al.}}(2018)}]{Najafi18}%
  \BibitemOpen
  \bibfield  {author} {\bibinfo {author} {\bibnamefont {{Najafi, J., et
  al.}}},\ }\bibfield  {title} {\enquote {\bibinfo {title} {Flagellar number
  governs bacterial spreading and transport efficiency},}\ }\href@noop {}
  {\bibfield  {journal} {\bibinfo  {journal} {Sci. Adv.}\ }\textbf {\bibinfo
  {volume} {4}},\ \bibinfo {pages} {eaar6425} (\bibinfo {year}
  {2018})}\BibitemShut {NoStop}%
\bibitem [{\citenamefont {Yang}\ \emph {et~al.}(2021)\citenamefont {Yang},
  \citenamefont {Zhu}, \citenamefont {Liu}, \citenamefont {Liu}, \citenamefont
  {Shi}, \citenamefont {Chen}, \citenamefont {Zheng}, \citenamefont {Ye},\ and\
  \citenamefont {Yang}}]{Yang21}%
  \BibitemOpen
  \bibfield  {author} {\bibinfo {author} {\bibfnamefont {Qing}\ \bibnamefont
  {Yang}}, \bibinfo {author} {\bibfnamefont {Hongwei}\ \bibnamefont {Zhu}},
  \bibinfo {author} {\bibfnamefont {Peng}\ \bibnamefont {Liu}}, \bibinfo
  {author} {\bibfnamefont {Rui}\ \bibnamefont {Liu}}, \bibinfo {author}
  {\bibfnamefont {Qingfan}\ \bibnamefont {Shi}}, \bibinfo {author}
  {\bibfnamefont {Ke}~\bibnamefont {Chen}}, \bibinfo {author} {\bibfnamefont
  {Ning}\ \bibnamefont {Zheng}}, \bibinfo {author} {\bibfnamefont {Fangfu}\
  \bibnamefont {Ye}}, \ and\ \bibinfo {author} {\bibfnamefont {Mingcheng}\
  \bibnamefont {Yang}},\ }\bibfield  {title} {\enquote {\bibinfo {title}
  {Topologically protected transport of cargo in a chiral active fluid aided by
  odd-viscosity-enhanced depletion interactions},}\ }\href@noop {} {\bibfield
  {journal} {\bibinfo  {journal} {Phys. Rev. Lett.}\ }\textbf {\bibinfo
  {volume} {126}},\ \bibinfo {pages} {198001} (\bibinfo {year}
  {2021})}\BibitemShut {NoStop}%
\bibitem [{\citenamefont {Aragones}\ \emph {et~al.}(2016)\citenamefont
  {Aragones}, \citenamefont {Steimel},\ and\ \citenamefont
  {Alexander-Katz}}]{Aragones16}%
  \BibitemOpen
  \bibfield  {author} {\bibinfo {author} {\bibfnamefont {J.~L.}\ \bibnamefont
  {Aragones}}, \bibinfo {author} {\bibfnamefont {J.~P.}\ \bibnamefont
  {Steimel}}, \ and\ \bibinfo {author} {\bibfnamefont {A.}~\bibnamefont
  {Alexander-Katz}},\ }\bibfield  {title} {\enquote {\bibinfo {title}
  {Elasticity-induced force reversal between active spinning particles in dense
  passive media},}\ }\href@noop {} {\bibfield  {journal} {\bibinfo  {journal}
  {Nat. Commun.}\ }\textbf {\bibinfo {volume} {7}},\ \bibinfo {pages} {11325}
  (\bibinfo {year} {2016})}\BibitemShut {NoStop}%
\bibitem [{\citenamefont {Aminov}\ \emph {et~al.}(2015)\citenamefont {Aminov},
  \citenamefont {Kafri},\ and\ \citenamefont {Kardar}}]{Aminov15}%
  \BibitemOpen
  \bibfield  {author} {\bibinfo {author} {\bibfnamefont {Avi}\ \bibnamefont
  {Aminov}}, \bibinfo {author} {\bibfnamefont {Yariv}\ \bibnamefont {Kafri}}, \
  and\ \bibinfo {author} {\bibfnamefont {Mehran}\ \bibnamefont {Kardar}},\
  }\bibfield  {title} {\enquote {\bibinfo {title} {Fluctuation-induced forces
  in nonequilibrium diffusive dynamics},}\ }\href@noop {} {\bibfield  {journal}
  {\bibinfo  {journal} {Phys. Rev. Lett.}\ }\textbf {\bibinfo {volume} {114}},\
  \bibinfo {pages} {230602} (\bibinfo {year} {2015})}\BibitemShut {NoStop}%
\bibitem [{\citenamefont {Lee}\ \emph {et~al.}(2017)\citenamefont {Lee},
  \citenamefont {Vella},\ and\ \citenamefont {Wettlaufer}}]{Lee17}%
  \BibitemOpen
  \bibfield  {author} {\bibinfo {author} {\bibfnamefont {Alpha~A.}\
  \bibnamefont {Lee}}, \bibinfo {author} {\bibfnamefont {Dominic}\ \bibnamefont
  {Vella}}, \ and\ \bibinfo {author} {\bibfnamefont {John~S.}\ \bibnamefont
  {Wettlaufer}},\ }\bibfield  {title} {\enquote {\bibinfo {title} {Fluctuation
  spectra and force generation in nonequilibrium systems},}\ }\href@noop {}
  {\bibfield  {journal} {\bibinfo  {journal} {Proc. Natl. Acad. Sci. USA}\
  }\textbf {\bibinfo {volume} {114}},\ \bibinfo {pages} {9255--9260} (\bibinfo
  {year} {2017})}\BibitemShut {NoStop}%
\bibitem [{\citenamefont {Liebchen}\ and\ \citenamefont
  {Levis}(2017)}]{Liebchen17}%
  \BibitemOpen
  \bibfield  {author} {\bibinfo {author} {\bibfnamefont {Benno}\ \bibnamefont
  {Liebchen}}\ and\ \bibinfo {author} {\bibfnamefont {Demian}\ \bibnamefont
  {Levis}},\ }\bibfield  {title} {\enquote {\bibinfo {title} {Collective
  behavior of chiral active matter: Pattern formation and enhanced flocking},}\
  }\href@noop {} {\bibfield  {journal} {\bibinfo  {journal} {Phys. Rev. Lett.}\
  }\textbf {\bibinfo {volume} {119}},\ \bibinfo {pages} {058002} (\bibinfo
  {year} {2017})}\BibitemShut {NoStop}%
\bibitem [{\citenamefont {Liebchen}\ and\ \citenamefont
  {Levis}(2022)}]{Liebchen22}%
  \BibitemOpen
  \bibfield  {author} {\bibinfo {author} {\bibfnamefont {Benno}\ \bibnamefont
  {Liebchen}}\ and\ \bibinfo {author} {\bibfnamefont {Demian}\ \bibnamefont
  {Levis}},\ }\bibfield  {title} {\enquote {\bibinfo {title} {Chiral active
  matter},}\ }\href@noop {} {\bibfield  {journal} {\bibinfo  {journal}
  {Europhys. Lett.}\ }\textbf {\bibinfo {volume} {139}},\ \bibinfo {pages}
  {67001} (\bibinfo {year} {2022})}\BibitemShut {NoStop}%
\bibitem [{\citenamefont {Shaebani}\ \emph {et~al.}(2012)\citenamefont
  {Shaebani}, \citenamefont {Sarabadani},\ and\ \citenamefont
  {Wolf}}]{Shaebani12}%
  \BibitemOpen
  \bibfield  {author} {\bibinfo {author} {\bibfnamefont {M.~Reza}\ \bibnamefont
  {Shaebani}}, \bibinfo {author} {\bibfnamefont {Jalal}\ \bibnamefont
  {Sarabadani}}, \ and\ \bibinfo {author} {\bibfnamefont {Dietrich~E.}\
  \bibnamefont {Wolf}},\ }\bibfield  {title} {\enquote {\bibinfo {title}
  {Nonadditivity of fluctuation-induced forces in fluidized granular media},}\
  }\href@noop {} {\bibfield  {journal} {\bibinfo  {journal} {Phys. Rev. Lett.}\
  }\textbf {\bibinfo {volume} {108}},\ \bibinfo {pages} {198001} (\bibinfo
  {year} {2012})}\BibitemShut {NoStop}%
\bibitem [{\citenamefont {Cattuto}\ \emph {et~al.}(2006)\citenamefont
  {Cattuto}, \citenamefont {Brito}, \citenamefont {Marconi}, \citenamefont
  {Nori},\ and\ \citenamefont {Soto}}]{Cattuto06}%
  \BibitemOpen
  \bibfield  {author} {\bibinfo {author} {\bibfnamefont {C.}~\bibnamefont
  {Cattuto}}, \bibinfo {author} {\bibfnamefont {R.}~\bibnamefont {Brito}},
  \bibinfo {author} {\bibfnamefont {U.~Marini~Bettolo}\ \bibnamefont
  {Marconi}}, \bibinfo {author} {\bibfnamefont {F.}~\bibnamefont {Nori}}, \
  and\ \bibinfo {author} {\bibfnamefont {R.}~\bibnamefont {Soto}},\ }\bibfield
  {title} {\enquote {\bibinfo {title} {Fluctuation-induced casimir forces in
  granular fluids},}\ }\href@noop {} {\bibfield  {journal} {\bibinfo  {journal}
  {Phys. Rev. Lett.}\ }\textbf {\bibinfo {volume} {96}},\ \bibinfo {pages}
  {178001} (\bibinfo {year} {2006})}\BibitemShut {NoStop}%
\bibitem [{\citenamefont {Shaebani}\ \emph {et~al.}(2013)\citenamefont
  {Shaebani}, \citenamefont {Sarabadani},\ and\ \citenamefont
  {Wolf}}]{Shaebani13}%
  \BibitemOpen
  \bibfield  {author} {\bibinfo {author} {\bibfnamefont {M.~Reza}\ \bibnamefont
  {Shaebani}}, \bibinfo {author} {\bibfnamefont {Jalal}\ \bibnamefont
  {Sarabadani}}, \ and\ \bibinfo {author} {\bibfnamefont {Dietrich~E.}\
  \bibnamefont {Wolf}},\ }\bibfield  {title} {\enquote {\bibinfo {title}
  {Long-range interactions in randomly driven granular fluids},}\ }\href@noop
  {} {\bibfield  {journal} {\bibinfo  {journal} {Phys. Rev. E}\ }\textbf
  {\bibinfo {volume} {88}},\ \bibinfo {pages} {022202} (\bibinfo {year}
  {2013})}\BibitemShut {NoStop}%
\bibitem [{\citenamefont {Ni}\ \emph {et~al.}(2015)\citenamefont {Ni},
  \citenamefont {Cohen~Stuart},\ and\ \citenamefont {Bolhuis}}]{Ni15}%
  \BibitemOpen
  \bibfield  {author} {\bibinfo {author} {\bibfnamefont {Ran}\ \bibnamefont
  {Ni}}, \bibinfo {author} {\bibfnamefont {Martien~A.}\ \bibnamefont
  {Cohen~Stuart}}, \ and\ \bibinfo {author} {\bibfnamefont {Peter~G.}\
  \bibnamefont {Bolhuis}},\ }\bibfield  {title} {\enquote {\bibinfo {title}
  {Tunable long range forces mediated by self-propelled colloidal hard
  spheres},}\ }\href@noop {} {\bibfield  {journal} {\bibinfo  {journal} {Phys.
  Rev. Lett.}\ }\textbf {\bibinfo {volume} {114}},\ \bibinfo {pages} {018302}
  (\bibinfo {year} {2015})}\BibitemShut {NoStop}%
\bibitem [{\citenamefont {Leite}\ \emph {et~al.}(2016)\citenamefont {Leite},
  \citenamefont {Lucena}, \citenamefont {Potiguar},\ and\ \citenamefont
  {Ferreira}}]{Leite16}%
  \BibitemOpen
  \bibfield  {author} {\bibinfo {author} {\bibfnamefont {L.~R.}\ \bibnamefont
  {Leite}}, \bibinfo {author} {\bibfnamefont {D.}~\bibnamefont {Lucena}},
  \bibinfo {author} {\bibfnamefont {F.~Q.}\ \bibnamefont {Potiguar}}, \ and\
  \bibinfo {author} {\bibfnamefont {W.~P.}\ \bibnamefont {Ferreira}},\
  }\bibfield  {title} {\enquote {\bibinfo {title} {Depletion forces on circular
  and elliptical obstacles induced by active matter},}\ }\href@noop {}
  {\bibfield  {journal} {\bibinfo  {journal} {Phys. Rev. E}\ }\textbf {\bibinfo
  {volume} {94}},\ \bibinfo {pages} {062602} (\bibinfo {year}
  {2016})}\BibitemShut {NoStop}%
\bibitem [{\citenamefont {Ray}\ \emph {et~al.}(2014)\citenamefont {Ray},
  \citenamefont {Reichhardt},\ and\ \citenamefont {Reichhardt}}]{Ray14}%
  \BibitemOpen
  \bibfield  {author} {\bibinfo {author} {\bibfnamefont {D.}~\bibnamefont
  {Ray}}, \bibinfo {author} {\bibfnamefont {C.}~\bibnamefont {Reichhardt}}, \
  and\ \bibinfo {author} {\bibfnamefont {C.~J.~Olson}\ \bibnamefont
  {Reichhardt}},\ }\bibfield  {title} {\enquote {\bibinfo {title} {Casimir
  effect in active matter systems},}\ }\href@noop {} {\bibfield  {journal}
  {\bibinfo  {journal} {Phys. Rev. E}\ }\textbf {\bibinfo {volume} {90}},\
  \bibinfo {pages} {013019} (\bibinfo {year} {2014})}\BibitemShut {NoStop}%
\bibitem [{\citenamefont {Feng}\ \emph {et~al.}(2021)\citenamefont {Feng},
  \citenamefont {Lei},\ and\ \citenamefont {Zhao}}]{Feng21}%
  \BibitemOpen
  \bibfield  {author} {\bibinfo {author} {\bibfnamefont {Fane}\ \bibnamefont
  {Feng}}, \bibinfo {author} {\bibfnamefont {Ting}\ \bibnamefont {Lei}}, \ and\
  \bibinfo {author} {\bibfnamefont {Nanrong}\ \bibnamefont {Zhao}},\ }\bibfield
   {title} {\enquote {\bibinfo {title} {Tunable depletion force in active and
  crowded environments},}\ }\href@noop {} {\bibfield  {journal} {\bibinfo
  {journal} {Phys. Rev. E}\ }\textbf {\bibinfo {volume} {103}},\ \bibinfo
  {pages} {022604} (\bibinfo {year} {2021})}\BibitemShut {NoStop}%
\bibitem [{\citenamefont {{Angelani, L., et al.}}(2011)}]{Angelani11}%
  \BibitemOpen
  \bibfield  {author} {\bibinfo {author} {\bibnamefont {{Angelani, L., et
  al.}}},\ }\bibfield  {title} {\enquote {\bibinfo {title} {Effective
  interactions between colloidal particles suspended in a bath of swimming
  cells},}\ }\href@noop {} {\bibfield  {journal} {\bibinfo  {journal} {Phys.
  Rev. Lett.}\ }\textbf {\bibinfo {volume} {107}},\ \bibinfo {pages} {138302}
  (\bibinfo {year} {2011})}\BibitemShut {NoStop}%
\bibitem [{\citenamefont {Liu}\ \emph {et~al.}(2020)\citenamefont {Liu},
  \citenamefont {Ye}, \citenamefont {Ye}, \citenamefont {Chen},\ and\
  \citenamefont {Yang}}]{Liu20}%
  \BibitemOpen
  \bibfield  {author} {\bibinfo {author} {\bibfnamefont {Peng}\ \bibnamefont
  {Liu}}, \bibinfo {author} {\bibfnamefont {Simin}\ \bibnamefont {Ye}},
  \bibinfo {author} {\bibfnamefont {Fangfu}\ \bibnamefont {Ye}}, \bibinfo
  {author} {\bibfnamefont {Ke}~\bibnamefont {Chen}}, \ and\ \bibinfo {author}
  {\bibfnamefont {Mingcheng}\ \bibnamefont {Yang}},\ }\bibfield  {title}
  {\enquote {\bibinfo {title} {Constraint dependence of active depletion forces
  on passive particles},}\ }\href@noop {} {\bibfield  {journal} {\bibinfo
  {journal} {Phys. Rev. Lett.}\ }\textbf {\bibinfo {volume} {124}},\ \bibinfo
  {pages} {158001} (\bibinfo {year} {2020})}\BibitemShut {NoStop}%
\bibitem [{\citenamefont {Harder}\ \emph {et~al.}(2014)\citenamefont {Harder},
  \citenamefont {Mallory}, \citenamefont {Tung}, \citenamefont {Valeriani},\
  and\ \citenamefont {Cacciuto}}]{Harder14}%
  \BibitemOpen
  \bibfield  {author} {\bibinfo {author} {\bibfnamefont {J.}~\bibnamefont
  {Harder}}, \bibinfo {author} {\bibfnamefont {S.~A.}\ \bibnamefont {Mallory}},
  \bibinfo {author} {\bibfnamefont {C.}~\bibnamefont {Tung}}, \bibinfo {author}
  {\bibfnamefont {C.}~\bibnamefont {Valeriani}}, \ and\ \bibinfo {author}
  {\bibfnamefont {A.}~\bibnamefont {Cacciuto}},\ }\bibfield  {title} {\enquote
  {\bibinfo {title} {The role of particle shape in active depletion},}\
  }\href@noop {} {\bibfield  {journal} {\bibinfo  {journal} {J. Chem. Phys.}\
  }\textbf {\bibinfo {volume} {141}},\ \bibinfo {pages} {194901} (\bibinfo
  {year} {2014})}\BibitemShut {NoStop}%
\bibitem [{\citenamefont {Baek}\ \emph {et~al.}(2018)\citenamefont {Baek},
  \citenamefont {Solon}, \citenamefont {Xu}, \citenamefont {Nikola},\ and\
  \citenamefont {Kafri}}]{Baek18}%
  \BibitemOpen
  \bibfield  {author} {\bibinfo {author} {\bibfnamefont {Yongjoo}\ \bibnamefont
  {Baek}}, \bibinfo {author} {\bibfnamefont {Alexandre~P.}\ \bibnamefont
  {Solon}}, \bibinfo {author} {\bibfnamefont {Xinpeng}\ \bibnamefont {Xu}},
  \bibinfo {author} {\bibfnamefont {Nikolai}\ \bibnamefont {Nikola}}, \ and\
  \bibinfo {author} {\bibfnamefont {Yariv}\ \bibnamefont {Kafri}},\ }\bibfield
  {title} {\enquote {\bibinfo {title} {Generic long-range interactions between
  passive bodies in an active fluid},}\ }\href@noop {} {\bibfield  {journal}
  {\bibinfo  {journal} {Phys. Rev. Lett.}\ }\textbf {\bibinfo {volume} {120}},\
  \bibinfo {pages} {058002} (\bibinfo {year} {2018})}\BibitemShut {NoStop}%
\bibitem [{\citenamefont {Parra-Rojas}\ and\ \citenamefont
  {Soto}(2014)}]{ParraRojas14}%
  \BibitemOpen
  \bibfield  {author} {\bibinfo {author} {\bibfnamefont {C.}~\bibnamefont
  {Parra-Rojas}}\ and\ \bibinfo {author} {\bibfnamefont {R.}~\bibnamefont
  {Soto}},\ }\bibfield  {title} {\enquote {\bibinfo {title} {Casimir effect in
  swimmer suspensions},}\ }\href@noop {} {\bibfield  {journal} {\bibinfo
  {journal} {Phys. Rev. E}\ }\textbf {\bibinfo {volume} {90}},\ \bibinfo
  {pages} {013024} (\bibinfo {year} {2014})}\BibitemShut {NoStop}%
\bibitem [{\citenamefont {Leonardo}\ \emph {et~al.}(2010)\citenamefont
  {Leonardo}, \citenamefont {Angelani}, \citenamefont {Dell\'Arciprete},
  \citenamefont {Ruocco}, \citenamefont {Iebba}, \citenamefont {Schippa},
  \citenamefont {Conte}, \citenamefont {Mecarini}, \citenamefont {Angelis},\
  and\ \citenamefont {Fabrizio}}]{DiLeonardo10}%
  \BibitemOpen
  \bibfield  {author} {\bibinfo {author} {\bibfnamefont {R.~Di}\ \bibnamefont
  {Leonardo}}, \bibinfo {author} {\bibfnamefont {L.}~\bibnamefont {Angelani}},
  \bibinfo {author} {\bibfnamefont {D.}~\bibnamefont {Dell\'Arciprete}},
  \bibinfo {author} {\bibfnamefont {G.}~\bibnamefont {Ruocco}}, \bibinfo
  {author} {\bibfnamefont {V.}~\bibnamefont {Iebba}}, \bibinfo {author}
  {\bibfnamefont {S.}~\bibnamefont {Schippa}}, \bibinfo {author} {\bibfnamefont
  {M.~P.}\ \bibnamefont {Conte}}, \bibinfo {author} {\bibfnamefont
  {F.}~\bibnamefont {Mecarini}}, \bibinfo {author} {\bibfnamefont {F.~De}\
  \bibnamefont {Angelis}}, \ and\ \bibinfo {author} {\bibfnamefont {E.~Di}\
  \bibnamefont {Fabrizio}},\ }\bibfield  {title} {\enquote {\bibinfo {title}
  {Bacterial ratchet motors},}\ }\href@noop {} {\bibfield  {journal} {\bibinfo
  {journal} {Proc. Natl. Acad. Sci. USA}\ }\textbf {\bibinfo {volume} {107}},\
  \bibinfo {pages} {9541--9545} (\bibinfo {year} {2010})}\BibitemShut {NoStop}%
\bibitem [{\citenamefont {{van Zuiden, B. C., et al.}}(2016)}]{vanZuiden16}%
  \BibitemOpen
  \bibfield  {author} {\bibinfo {author} {\bibnamefont {{van Zuiden, B. C., et
  al.}}},\ }\bibfield  {title} {\enquote {\bibinfo {title} {Spatiotemporal
  order and emergent edge currents in active spinner materials},}\ }\href@noop
  {} {\bibfield  {journal} {\bibinfo  {journal} {Proc. Natl. Acad. Sci. USA}\
  }\textbf {\bibinfo {volume} {113}},\ \bibinfo {pages} {12919--12924}
  (\bibinfo {year} {2016})}\BibitemShut {NoStop}%
\bibitem [{\citenamefont {Tan}\ \emph {et~al.}(2022)\citenamefont {Tan},
  \citenamefont {Mietke}, \citenamefont {Li}, \citenamefont {Chen},
  \citenamefont {Higinbotham}, \citenamefont {Foster}, \citenamefont {Gokhale},
  \citenamefont {Dunkel},\ and\ \citenamefont {Fakhri}}]{Tan22}%
  \BibitemOpen
  \bibfield  {author} {\bibinfo {author} {\bibfnamefont {Tzer~Han}\
  \bibnamefont {Tan}}, \bibinfo {author} {\bibfnamefont {Alexander}\
  \bibnamefont {Mietke}}, \bibinfo {author} {\bibfnamefont {Junang}\
  \bibnamefont {Li}}, \bibinfo {author} {\bibfnamefont {Yuchao}\ \bibnamefont
  {Chen}}, \bibinfo {author} {\bibfnamefont {Hugh}\ \bibnamefont
  {Higinbotham}}, \bibinfo {author} {\bibfnamefont {Peter~J.}\ \bibnamefont
  {Foster}}, \bibinfo {author} {\bibfnamefont {Shreyas}\ \bibnamefont
  {Gokhale}}, \bibinfo {author} {\bibfnamefont {J{\"o}rn}\ \bibnamefont
  {Dunkel}}, \ and\ \bibinfo {author} {\bibfnamefont {Nikta}\ \bibnamefont
  {Fakhri}},\ }\bibfield  {title} {\enquote {\bibinfo {title} {Odd dynamics of
  living chiral crystals},}\ }\href@noop {} {\bibfield  {journal} {\bibinfo
  {journal} {Nature}\ }\textbf {\bibinfo {volume} {607}},\ \bibinfo {pages}
  {287--293} (\bibinfo {year} {2022})}\BibitemShut {NoStop}%
\bibitem [{\citenamefont {Petroff}\ \emph {et~al.}(2015)\citenamefont
  {Petroff}, \citenamefont {Wu},\ and\ \citenamefont {Libchaber}}]{Petroff15}%
  \BibitemOpen
  \bibfield  {author} {\bibinfo {author} {\bibfnamefont {Alexander~P.}\
  \bibnamefont {Petroff}}, \bibinfo {author} {\bibfnamefont {Xiao-Lun}\
  \bibnamefont {Wu}}, \ and\ \bibinfo {author} {\bibfnamefont {Albert}\
  \bibnamefont {Libchaber}},\ }\bibfield  {title} {\enquote {\bibinfo {title}
  {Fast-moving bacteria self-organize into active two-dimensional crystals of
  rotating cells},}\ }\href@noop {} {\bibfield  {journal} {\bibinfo  {journal}
  {Phys. Rev. Lett.}\ }\textbf {\bibinfo {volume} {114}},\ \bibinfo {pages}
  {158102} (\bibinfo {year} {2015})}\BibitemShut {NoStop}%
\bibitem [{\citenamefont {Weeks}\ \emph {et~al.}(1971)\citenamefont {Weeks},
  \citenamefont {Chandler},\ and\ \citenamefont {Andersen}}]{Weeks71}%
  \BibitemOpen
  \bibfield  {author} {\bibinfo {author} {\bibfnamefont {John~D.}\ \bibnamefont
  {Weeks}}, \bibinfo {author} {\bibfnamefont {David}\ \bibnamefont {Chandler}},
  \ and\ \bibinfo {author} {\bibfnamefont {Hans~C.}\ \bibnamefont {Andersen}},\
  }\bibfield  {title} {\enquote {\bibinfo {title} {Role of repulsive forces in
  determining the equilibrium structure of simple liquids},}\ }\href@noop {}
  {\bibfield  {journal} {\bibinfo  {journal} {J. Chem. Phys.}\ }\textbf
  {\bibinfo {volume} {54}},\ \bibinfo {pages} {5237--5247} (\bibinfo {year}
  {1971})}\BibitemShut {NoStop}%
\bibitem [{\citenamefont {Nossal}\ and\ \citenamefont
  {Weiss}(1974)}]{Nossal74}%
  \BibitemOpen
  \bibfield  {author} {\bibinfo {author} {\bibfnamefont {Ralph}\ \bibnamefont
  {Nossal}}\ and\ \bibinfo {author} {\bibfnamefont {George~H.}\ \bibnamefont
  {Weiss}},\ }\bibfield  {title} {\enquote {\bibinfo {title} {A descriptive
  theory of cell migration on surfaces},}\ }\href@noop {} {\bibfield  {journal}
  {\bibinfo  {journal} {J. Theor. Biol.}\ }\textbf {\bibinfo {volume} {47}},\
  \bibinfo {pages} {103--113} (\bibinfo {year} {1974})}\BibitemShut {NoStop}%
\bibitem [{\citenamefont {Shaebani}\ and\ \citenamefont
  {Rieger}(2019)}]{Shaebani19}%
  \BibitemOpen
  \bibfield  {author} {\bibinfo {author} {\bibfnamefont {M.~Reza}\ \bibnamefont
  {Shaebani}}\ and\ \bibinfo {author} {\bibfnamefont {Heiko}\ \bibnamefont
  {Rieger}},\ }\bibfield  {title} {\enquote {\bibinfo {title} {Transient
  anomalous diffusion in run-and-tumble dynamics},}\ }\href@noop {} {\bibfield
  {journal} {\bibinfo  {journal} {Front. Phys.}\ }\textbf {\bibinfo {volume}
  {7}},\ \bibinfo {pages} {120} (\bibinfo {year} {2019})}\BibitemShut {NoStop}%
\bibitem [{\citenamefont {Shaebani}\ \emph
  {et~al.}(2022{\natexlab{b}})\citenamefont {Shaebani}, \citenamefont
  {Rieger},\ and\ \citenamefont {Sadjadi}}]{Shaebani22b}%
  \BibitemOpen
  \bibfield  {author} {\bibinfo {author} {\bibfnamefont {M.~Reza}\ \bibnamefont
  {Shaebani}}, \bibinfo {author} {\bibfnamefont {Heiko}\ \bibnamefont
  {Rieger}}, \ and\ \bibinfo {author} {\bibfnamefont {Zeinab}\ \bibnamefont
  {Sadjadi}},\ }\bibfield  {title} {\enquote {\bibinfo {title} {Kinematics of
  persistent random walkers with two distinct modes of motion},}\ }\href@noop
  {} {\bibfield  {journal} {\bibinfo  {journal} {Phys. Rev. E}\ }\textbf
  {\bibinfo {volume} {106}},\ \bibinfo {pages} {034105} (\bibinfo {year}
  {2022}{\natexlab{b}})}\BibitemShut {NoStop}%
\bibitem [{\citenamefont {Wioland}\ \emph {et~al.}(2013)\citenamefont
  {Wioland}, \citenamefont {Woodhouse}, \citenamefont {Dunkel}, \citenamefont
  {Kessler},\ and\ \citenamefont {Goldstein}}]{Wioland13}%
  \BibitemOpen
  \bibfield  {author} {\bibinfo {author} {\bibfnamefont {Hugo}\ \bibnamefont
  {Wioland}}, \bibinfo {author} {\bibfnamefont {Francis~G.}\ \bibnamefont
  {Woodhouse}}, \bibinfo {author} {\bibfnamefont {J\"orn}\ \bibnamefont
  {Dunkel}}, \bibinfo {author} {\bibfnamefont {John~O.}\ \bibnamefont
  {Kessler}}, \ and\ \bibinfo {author} {\bibfnamefont {Raymond~E.}\
  \bibnamefont {Goldstein}},\ }\bibfield  {title} {\enquote {\bibinfo {title}
  {Confinement stabilizes a bacterial suspension into a spiral vortex},}\
  }\href@noop {} {\bibfield  {journal} {\bibinfo  {journal} {Phys. Rev. Lett.}\
  }\textbf {\bibinfo {volume} {110}},\ \bibinfo {pages} {268102} (\bibinfo
  {year} {2013})}\BibitemShut {NoStop}%
\bibitem [{\citenamefont {Sadjadi}\ and\ \citenamefont
  {Shaebani}(2021)}]{Sadjadi21}%
  \BibitemOpen
  \bibfield  {author} {\bibinfo {author} {\bibfnamefont {Zeinab}\ \bibnamefont
  {Sadjadi}}\ and\ \bibinfo {author} {\bibfnamefont {M.~Reza}\ \bibnamefont
  {Shaebani}},\ }\bibfield  {title} {\enquote {\bibinfo {title} {Orientational
  memory of active particles in multistate non-markovian processes},}\
  }\href@noop {} {\bibfield  {journal} {\bibinfo  {journal} {Phys. Rev. E}\
  }\textbf {\bibinfo {volume} {104}},\ \bibinfo {pages} {054613} (\bibinfo
  {year} {2021})}\BibitemShut {NoStop}%
\bibitem [{\citenamefont {Lei}\ \emph {et~al.}(2019)\citenamefont {Lei},
  \citenamefont {Ciamarra},\ and\ \citenamefont {Ni}}]{Lei19}%
  \BibitemOpen
  \bibfield  {author} {\bibinfo {author} {\bibfnamefont {Qun-Li}\ \bibnamefont
  {Lei}}, \bibinfo {author} {\bibfnamefont {Massimo~Pica}\ \bibnamefont
  {Ciamarra}}, \ and\ \bibinfo {author} {\bibfnamefont {Ran}\ \bibnamefont
  {Ni}},\ }\bibfield  {title} {\enquote {\bibinfo {title} {Nonequilibrium
  strongly hyperuniform fluids of circle active particles with large local
  density fluctuations},}\ }\href@noop {} {\bibfield  {journal} {\bibinfo
  {journal} {Sci. Adv.}\ }\textbf {\bibinfo {volume} {5}},\ \bibinfo {pages}
  {eaau7423} (\bibinfo {year} {2019})}\BibitemShut {NoStop}%
\bibitem [{\citenamefont {Zhang}\ and\ \citenamefont
  {Snezhko}(2022)}]{Zhang22}%
  \BibitemOpen
  \bibfield  {author} {\bibinfo {author} {\bibfnamefont {Bo}~\bibnamefont
  {Zhang}}\ and\ \bibinfo {author} {\bibfnamefont {Alexey}\ \bibnamefont
  {Snezhko}},\ }\bibfield  {title} {\enquote {\bibinfo {title} {Hyperuniform
  active chiral fluids with tunable internal structure},}\ }\href@noop {}
  {\bibfield  {journal} {\bibinfo  {journal} {Phys. Rev. Lett.}\ }\textbf
  {\bibinfo {volume} {128}},\ \bibinfo {pages} {218002} (\bibinfo {year}
  {2022})}\BibitemShut {NoStop}%
\bibitem [{\citenamefont {Torrik}\ \emph {et~al.}(2021)\citenamefont {Torrik},
  \citenamefont {Naji},\ and\ \citenamefont {Zarif}}]{Torrik21}%
  \BibitemOpen
  \bibfield  {author} {\bibinfo {author} {\bibfnamefont {Abdolhalim}\
  \bibnamefont {Torrik}}, \bibinfo {author} {\bibfnamefont {Ali}\ \bibnamefont
  {Naji}}, \ and\ \bibinfo {author} {\bibfnamefont {Mahdi}\ \bibnamefont
  {Zarif}},\ }\bibfield  {title} {\enquote {\bibinfo {title} {Dimeric colloidal
  inclusion in a chiral active bath: Effective interactions and
  chirality-induced torque},}\ }\href@noop {} {\bibfield  {journal} {\bibinfo
  {journal} {Phys. Rev. E}\ }\textbf {\bibinfo {volume} {104}},\ \bibinfo
  {pages} {064610} (\bibinfo {year} {2021})}\BibitemShut {NoStop}%
\bibitem [{LAM(LAMMPS molecular dynamics simulator)}]{LAMMPS}%
  \BibitemOpen
  \href@noop {} {\enquote {\bibinfo {title} {http://lammps.sandia.gov/},}\ }
  (\bibinfo {year} {LAMMPS molecular dynamics simulator})\BibitemShut {NoStop}%
\bibitem [{\citenamefont {Plimpton}(1995)}]{Plimpton95}%
  \BibitemOpen
  \bibfield  {author} {\bibinfo {author} {\bibfnamefont {Steve}\ \bibnamefont
  {Plimpton}},\ }\bibfield  {title} {\enquote {\bibinfo {title} {Fast parallel
  algorithms for short-range molecular dynamics},}\ }\href@noop {} {\bibfield
  {journal} {\bibinfo  {journal} {J. Comput. Phys.}\ }\textbf {\bibinfo
  {volume} {117}},\ \bibinfo {pages} {1--19} (\bibinfo {year}
  {1995})}\BibitemShut {NoStop}%
\end{thebibliography}%

\end{document}